\documentclass[reprint,eqsecnum,floats,aps,amsmath,amssymb,nofootinbib,prd,onecolumn,showpacs,superscriptaddress]{revtex4-2}

\usepackage{graphicx}
\usepackage{bm}
\usepackage{amsmath}	
\usepackage{amsfonts}
\usepackage{mathtools}
\usepackage{amssymb}							
\usepackage{braket}
\usepackage{amsthm}			
\usepackage{hyperref}							

\usepackage[normalem]{ulem}
\usepackage[T1]{fontenc} 

\usepackage{color}
\usepackage[dvipsnames]{xcolor}

\usepackage{nomencl}
\makenomenclature

\begin{document}

\title{Parametrization of the primordial power spectrum in loop quantum cosmology}

\author{Almudena Guill\'en}
\email{almudena.guillen@iem.cfmac.csic.es}
\affiliation{Instituto de Estructura de la Materia, IEM-CSIC, Serrano 121, 28006 Madrid, Spain}

\author{Kai Langer}
\email{kai.langer@fau.de}
\affiliation{Institute for Quantum Gravity, Theoretical Physics III, Department of Physics,  Friedrich-Alexander-Universit\"at Erlangen-N\"urnberg, Staudtstr. 7, 91058 Erlangen, Germany.}

\author{Guillermo A. Mena Marug\'an}
\email{mena@iem.cfmac.csic.es}
\affiliation{Instituto de Estructura de la Materia, IEM-CSIC, Serrano 121, 28006 Madrid, Spain}

\author{Niklas Rodenbücher}
\email{niklas.rodenbuecher@fau.de}
\affiliation{Institute for Quantum Gravity, Theoretical Physics III, Department of Physics,  Friedrich-Alexander-Universit\"at Erlangen-N\"urnberg, Staudtstr. 7, 91058 Erlangen, Germany.}

\author{Antonio Vicente-Becerril}
\email{antonio.vicenteb@estudiante.uam.es}
\affiliation{Instituto de Estructura de la Materia, IEM-CSIC, Serrano 121, 28006 Madrid, Spain}

\begin{abstract}
We investigate the imprints on the angular power spectra of cosmological perturbations of a pre-inflationary bounce phase, as described by the hybrid and dressed metric approaches to loop quantum cosmology. For this purpose, we derive a new parametrization of the primordial power spectrum at the end of the inflationary regime. Apart from slow-roll coefficients and cosmological parameters that are present in the standard cosmological scenario without quantum modifications, this parametrization additionally depends only on pre-inflationary physics. More specifically, we find a dependence on the number of e-folds during the bounce epoch and on a characteristic suppression scale which, given the e-folds accumulated during cosmic evolution, is determined by the energy density at the bounce. Recall that this density depends on the Immirzi parameter and the area gap known from LQG. This leads to a robust and accurate parametrization of the primordial power spectrum. Since in pre-inflationary scenarios there is no preferred vacuum state, we adopt the NO-AHD proposal, which selects a vacuum that is optimally adapted to the background dynamics and yields a non-oscillatory primordial power spectrum. With this choice, we show that the tensor-to-scalar ratio in both quantization approaches coincides with its expression in the standard $\Lambda$CDM model when the observed scales are not much smaller than the power-suppressed region. Computing also the angular power spectrum, we find that, for a total cosmic expansion of about 140 e-folds, both the hybrid and the dressed metric approaches exhibit excellent agreement with Planck data at high multipoles, while apparently improving the fit with respect to $\Lambda$CDM for low multipole numbers.
\end{abstract}

\maketitle

\section{Introduction}
\label{sec:intro}

The discovery of the cosmic microwave background (CMB) \cite{CMB_detection} marked a milestone in our understanding of the early universe. In recent years, precise measurements of the CMB by collaborations such as WMAP and Planck have made it possible to confront theoretical predictions with observations with remarkable precision \cite{WMAP,Planck_anomalies,Planck_parameters,Planck_Constraints}. In this context, the $\Lambda$CDM model
(where $\Lambda$ is the cosmological constant and CDM stands for the initials of cold dark matter)
has gained significant credibility, as its predictions are in good agreement with the observed power spectra of CMB temperature and polarization (as well as with other cosmological probes such as fluctuations in the late-time density field \cite{DESY3, KiDS_legacy}). However, these data have also revealed the existence of of certain tensions and possible anomalies within the standard $\Lambda$CDM model. This is the case, for instance, of the power suppression anomaly, the lensing amplitude anomaly and the parity anomaly \cite{Planck_anomalies,ASr,ASr2,AgSr,AgSr2}. It has been suggested that such tensions may indicate the presence of new physics, arising from a non-standard inflationary evolution, from pre-inflationary epochs, or even from alternative processes to inflation \cite{Singh1}. A possible mechanism to (at least) alleviate these tensions is the introduction of a power suppression at large scales, characterized by a suppression scale, sometimes also called a cutoff  \cite{cutoff1,cutoff2,cutoff3,cutoff4}. This mechanism could result from quantum gravity effects in the very early stages of our universe.

In this context, among other approaches loop quantum gravity (LQG) has emerged as a promising candidate for a quantum formulation of general relativity (GR). LQG is a non-perturbative quantization program for GR, originally based on the canonical formulation of Einstein theory in terms of the Ashtekar–Barbero variables \cite{LQG, Thie}. These real variables consist of an SU(2) gauge connection and a densitized triad, which capture the physical information about the spatial geometry and its extrinsic curvature. The application of LQG techniques to symmetry reduced models such as cosmology, aimed at studying quantum gravitational phenomena in the early universe and in astrophysical contexts, has led to the development of a discipline known as Loop Quantum Cosmology (LQC) \cite{LQC}. For various models within LQC a resolution of the cosmic singularity problem is achieved, replacing the conventional concept of the {\sl {Big Bang}} with a cosmic bounce, commonly referred to as the {\sl Big Bounce} \cite{APS,APS1}. This bounce represents a quantum transition that enables the universe to evolve from a contracting phase to an expanding one. Furthermore, LQC  models provide an explicit explanation for the emergence of power suppression at large scales, leading to an effective (not hard) cutoff in the primordial power spectrum (PPS) \cite{revAWW,hybrid_rev}. 

Several quantization approaches have been developed for the study of cosmological perturbations in LQC \cite{dressed1,dressed2,hybr_inf1,hybrid_rev,effective2,effective3,effective4,effective5}.
Two of these approaches respect the ultraviolet behavior of the cosmological approximations, preserving the dispersion relations and the hyperbolicity of the propagation equations for modes in that sector. These are the hybrid approach \cite{hybridCMB,hybr_inf1,hybrid_rev,hybr_inf2,hybr_ten} and the dressed metric approach \cite{AAN,AM,dressed1,dressed2,dressed4}. In this work, we will focus our attention on these two approaches. 

The quantization procedure is significantly different in these two approaches. In the hybrid case, the perturbations and the background are treated as a single constrained canonical system \cite{hybrid_rev}. The resulting Hamiltonian constraint is quantized by combining a loop quantization of the background with a conventional Fock quantization of the perturbations. Evolution equations for the linear perturbations (e.g. the Mukhanov-Sasaki equation \cite{Mukhanov,Sasaki,Sasaki1}) can then be derived from this constraint using a kind of mean-field approximation. For quantum states of the background that are sharply peaked on an effective LQC trajectory \cite{APS1,Taveras}, these equations reproduce those of GR, except for modifications in the background-dependent mass term in the Mukhanov-Sasaki equation, which is now obtained by evaluating the canonical background variables on the effective solution corresponding to the peak \cite{hybrid_rev,NBMmass}, yielding a modified mass term determined by the LQC background model. 

In contrast, in the dressed metric approach, one first quantizes the background geometry, considers sharply peaked states, assigns to their peaks a metric dressed with quantum corrections, and lifts this metric to the phase space of the linear perturbations \cite{AAN,dressed2}. Likewise to the hybrid approach the resulting propagation equations for the linear perturbations resemble those of GR. However, the background geometry differs from its GR counterpart, leading to a distinct modified background-dependent mass, in general different from the hybrid approach one,  especially near the bounce.

To evolve the cosmological perturbations through the cosmic expansion it is necessary to fix some initial conditions on them. In cosmology, the choice of the initial state of the perturbations is equivalent to the choice of a vacuum state. In standard inflationary cosmology, the Bunch–Davies vacuum is the most common choice, as it is invariant under the symmetries of de Sitter spacetime and is a Hadamard state \cite{Bunch,Mukhanov1,WaldH}. However, in LQC scenarios, this vacuum is not expected to be the obvious choice, since there exist modes with wavelengths of the order of the Planck scale that experience quantum effects from the pre-inflationary stages near the bounce \cite{NM}. In such cases, selecting a preferred vacuum becomes nontrivial, as in generic non-stationary curved spacetimes there is no preferred vacuum based on symmetry invariance. In this work, we adopt the NO-AHD (Non-Oscillatory Adiabatic Hamiltonian Diagonalization) proposal \cite{NMT,NMP} for the choice of a vacuum. The proposal is based on a diagonalization of the Hamiltonian of the perturbations in the asymptotic limit of infinitely large wave numbers (ultraviolet modes), which allows for a natural choice of positive-frequency modes in the ultraviolet sector \cite{NMT}. This proposed vacuum state is specially well adapted to the background dynamics, and leads to a power spectrum without rapid oscillations, thereby suppressing spurious power contributions that could arise from averaging over such oscillatory behavior \cite{NM, NMT,NMP}.  

We want to confront the predictions of the hybrid and dressed metric quantization approaches in LQC scenarios with CMB observations, in particular through the (temperature) angular power spectrum. For this purpose, we propose in this work a novel parametrization of the PPS at the end of inflation. Recall that the PPS, together with standard cosmological parameters, determines the angular power spectrum. Our parametrization explicitly exhibits a direct dependence on the quantum bounce period and on the choice of the vacuum state. More precisely, we find that, in addition to the usual parameters found in the standard $\Lambda$CDM model with slow-roll inflation, our parametrization depends only on two additional quantities: the number of e-folds experienced during the quantum bounce period and the value of a suppression scale. Moreover, this scale is directly determined by the total number of e-folds accumulated during cosmic evolution from the bounce to the present day and the energy density at the bounce, related in turn to the Immirzi parameter and the area gap in LQG. We arrive at this parametrization by simplifying the expressions of the PPS for each of the (scalar or tensor) gauge invariant modes, comparing the different scales that play a role in the general form of this power spectrum. Our approximate analytic results show an extremely good agreement with numerical computations around the suppression scale and in the ultraviolet sector. This agreement is found for both the scalar and tensor perturbations in both of the two quantization approaches (hybrid and dressed metric) studied here.

Furthermore, we show that the deduced PPS reproduces the standard expression of the tensor-to-scalar ratio given by the slow-roll approximation for scales that are not much smaller that the suppression scale caused by LQC effects. This condition is in fact necessary for the typical pivot scale employed for CMB data if the predicted PPS must satisfactorily fit the observations, because no relevant suppression is observed around it. Finally, from the parametrized PPS, we compute the corresponding TT (temperature) angular power spectrum, showing a significant power decrease at low multipoles when the suppression scale lies within the observable window of wave numbers and the pivot scale is located in a region with a locally quasi-scale-invariant PPS. This suppression may provide a better fit to Planck observations than the standard $\Lambda$CDM model and potentially alleviate, at least, the low-$\ell$ anomaly in the TT-power spectrum.

In our analysis, we focus our attention on background solutions for which there may exist quantum effects on the modes observable in the CMB. These solutions with phenomenological interest should not experience a long inflationary stage, which would lead to a large number of e-folds that would completely erase the potential LQC effects in the PPS. In addition, they must not present an excessively short inflationary stage, producing a small number of e-folds, since this would be in tension with observations that bound this number from below \cite{efolds} and, furthermore, would generate an excessive power suppression at high multipoles, resulting in incompatibilities with Planck data.

The article is structured as follows: In Sec. II, we briefly review the dynamical equations of the effective LQC background and present the main results of the hybrid and dressed metric quantization approaches for scalar and tensor perturbations. Afterwards in Sec. III, we derive the scalar and tensor PPS and we parametrize these spectra at the end of the inflationary stage. We then analyze the agreement of this parametrization with numerical computations and discuss the implications for the PPS. Given these results, next in Sec. IV we compute the corresponding TT angular power spectrum and compare it with Planck observations. Our results demonstrate that the TT angular power spectrum has more power suppression at low multipoles than the best fit in the $\Lambda$CDM model. Finally, in Sec. V we summarize our results, provide conclusions and possible directions for future work. An appendix is added to make the article self-contained, with detailed formulas about the solutions of the dynamical background equations as well as the analytic solutions for the modes of the Mukhanov-Sasaki equation. 

Throughout the paper, we use Planck units, setting $G=c=\hbar=1$. We only employ other units when computing the angular power spectrum in order to confront the results with Planck data in Sec.\ V.

\section{The Framework}

In this section, we summarize the most relevant results concerning the dynamics of the effective background in LQC. We also present the main results for the tensor and scalar modes. For further technical details, we refer the reader to Appendix \ref{Appendix_A} and to the work in Refs. \cite{NM,NMY,AMV,MVY,MVY2}.

\subsection{Background: Effective flat FLRW equations}

We consider a homogeneous and isotropic background described by a spatially flat Friedmann-Lemaître-Robertson-Walker metric, characterized by a scale factor $a(t)$, in the presence of a homogeneous scalar inflaton field $\phi(t)$ with a potential $V(\phi)$. Exclusively for numerical computations and in regimes where the influence of this potential is not very relevant, we will take a quadratic potential as a reference example, with $V(\phi)= m^2\phi^2/2$ and $m=1.2\times 10^{-6}$ in Planck units to ensure consistency and facilitate comparison with previous results in the literature  \cite{AM,NM}. 

The background dynamics in effective LQC is governed by the following modified FLRW equations \cite{AAN,hybridCMB}
\begin{eqnarray}\label{eq_Friedman_LQC}
    \left( \frac{a'}{a}\right)^{2} = \frac{8\pi}{3}a^{2}\rho \left(1 - \frac{\rho}{\rho_{c}} \right),\quad \frac{a''}{a} = \frac{4\pi}{3}a^{2}\rho \left(1 + 2\frac{\rho}{\rho_{c}} \right) - 4\pi a^{2}P \left(1 - 2\frac{\rho}{\rho_{c}} \right),
\end{eqnarray}
where the prime denotes the derivative with respect to conformal time, and the energy density and pressure of the scalar field are $\rho = V(\phi)+ (\phi^{\prime})^2/(2a^2)$ and $P =\rho - 2 V(\phi)$, respectively. The inflaton reaches its maximum energy density at the bounce. This maximum is called the critical density, $\rho_c$. With the standard regularization of the Hamiltonian constraint used in LQC \cite{APS}, the corresponding value of this density is $\rho_c= 3 /(8\pi \gamma^2 \Delta)$ \cite{LQC}, where $\gamma$ is the Immirzi parameter \cite{immirzi} and $\Delta = 4 \sqrt{3}\pi \gamma $ is the area gap, obtained from the area spectrum of LQG \cite{LQG,Thie,ALArea}. Nevertheless, other suggested regularizations in LQC lead to slightly different values of this critical density \cite{Yang:2009fp}. To account for these different possibilities and also to explore whether the value of $\rho_c$ may be determined by observations, in the following we will treat this quantity as a parameter of the model under consideration. For concreteness, in the numerical computations carried out in this work we adopt the standard value of $\rho_c$ for $\gamma=0.2375$, which is a frequent choice of $\gamma$ motivated by certain black-hole entropy calculations in spherically symmetric reduced models in LQG \cite{LQG}. Moreover, this choice facilitates comparison with previous studies of the PPS in LQC available in the literature, where the same value of $\gamma$ was adopted (see e.g. \cite{ASr2,hybridCMB,AAN}). 

The background equations do not admit an analytic solution that remains valid throughout the entire cosmic evolution. However, in the solutions of phenomenological interest on which we focus our discussion, the contribution of the inflaton potential to its energy density is negligible approximately until the onset of inflation. As a consequence, the corresponding background dynamics can in fact be divided into three distinct epochs, each of which allows for an analytic description under the assumption of instantaneous transitions between them. The first epoch corresponds to the era just after the quantum bounce, during which the kinetically dominated evolution presents important quantum corrections. It is followed by a purely kinetic classical epoch, where the low energy density reached at the end of the bounce renders quantum corrections negligible, but the potential term is still not relevant and can still be neglected\footnote{In fact, one can invoke Hubble anti-friction in the Klein–Gordon equation during the contracting phase before the bounce to argue that the era immediately following this quantum bounce should be kinetically dominated  \cite{Singh_2004}.}. Finally, this kinetic phase evolves into an inflationary epoch, which is well described by the slow-roll approximation.

In the quantum bounce period, with a negligible potential contribution, the scale factor and the scalar field can be expressed in proper time as \cite{NM} 
\begin{equation}\label{scale_factor_LQC}
 a(t)=[1+24\pi \rho_c t^2]^{1/6}, \quad \text{and} \quad \dot{\phi }(t)=\pm \sqrt{2\rho_c}a^{-3}(t).
\end{equation}
Here and in the following, we set the cosmic time at the bounce to zero for simplicity. The relation between proper and conformal times is given by 
\begin{equation}
\label{confytime}
\eta={}_{2}F_1\left[ \frac{1}{6},\frac{1}{2};\frac{3}{2};-24\pi \rho_c t^2 \right]\,t,
\end{equation} 
where we have chosen to also set the conformal time to zero at the bounce, and $_{2}F_1$ is the Gaussian hypergeometric function \cite{NM}. Note that we have set the value of the scale factor at the bounce equal to one, $a_B=1$. Hence, in the following, the value of the scale factor at any other point can be reinterpreted as the ratio with respect to the value at the bounce, or equivalently as the exponential of the number of e-folds in the elapsed interval. 

As the scale factor increases with time, the energy density decreases rapidly,  allowing us to neglect quantum contributions in the background equations \eqref{eq_Friedman_LQC}, thereby leading to classical standard cosmological evolution equations. The scale factor and the Hubble parameter during this era can be written as
\begin{equation} \label{scale_factor_kinetic}
    a(\eta)=a_0\sqrt{1+2a_0H_0\,(\eta -\eta _0)}, \qquad H(\eta)= \left(\frac{a_0}{a(\eta)}\right)^3 H_0 ,
\end{equation}
where $a_0$  and $H_0$ are respectively the scale factor and the Hubble parameter at the end of the bounce epoch, defined at a proper time $t_0$, for which the conformal time is $\eta_0$ (see subsection \ref{sec:Perturbations} for more details\footnote{For our choice of $t_0$ in our approximations and computations, the resulting conformal time is $\eta_0 \simeq 0.35$ in the hybrid case and $\eta_0\simeq 0.63$ in the dressed metric case.}). 

The end of the kinetic period, on the other hand, is marked at the conformal time $\eta_i\simeq 912$ as determined by numerical background computations for typical solutions of phenomenological interest (for a quadratic inflaton potential) \cite{MVY,MVY2}. After this time, the potential energy of the inflaton surpasses its kinetic energy, giving rise to the slow-roll inflationary phase. For the subsequent analysis, we introduce the slow-roll parameters \cite{Langlois, Baumann}
\begin{equation}
\label{eq:SlowRoll}
   \epsilon _V=\frac{1}{16\pi }\left (\frac{V_{,\phi}(\phi )}{V(\phi )}  \right )^2, \qquad \delta _V=\frac{1}{8\pi} \frac{V_{,\phi \phi}(\phi )}{V(\phi )},
\end{equation}
where the comma in the potential denotes a derivative with respect to the inflaton field. We define the end of the slow-roll regime as the time when these parameters cease to be smaller than one, denoting that conformal time by $\eta_e\simeq 2788$, determined also by numerical background computations of typical solutions with phenomenological interest \cite{MVY,MVY2}. 
During slow roll, the dynamics of the background can be approximated by \cite{Langlois, Baumann} 
\begin{eqnarray}
       \frac{\mathrm{d} }{\mathrm{d} \eta }\biggl ( \frac{a}{a'} \biggl )\approx \epsilon _V-1,
\end{eqnarray}
which yields 
\begin{equation} \label{scale_factor_SR}
   \eta_e-\eta\approx  \frac{1}{aH}(1+\epsilon_V).
\end{equation}

\subsection{Linear Perturbations}
\label{sec:Perturbations}
In the two approaches to cosmological perturbations that we will consider in this work, namely the hybrid and dressed metric approach to cosmological perturbations in LQC, the evolution equations for the Mukhanov-Sasaki and tensor modes take the following form \cite{hybr_ten,NBMmass,LGomar}
\begin{equation}
\mu_{k}^{(t,s) \prime \prime}+\left[k^2+\mathbf{s}^{(t,s)}\right] \mu_{k}^{(t,s)}=0,    
\end{equation} where we denoted tensor and scalar modes by $t$ and $s$ respectively. The background-dependent mass term $\mathbf{s}^{(t,s)}$ differs in the hybrid and dressed metric approaches. We denote the hybrid version by $s^{(t,s)}$ and the dressed metric one by $\bar{s}^{(t,s)}$. Their explicit forms for the scalar and tensor modes are discussed in the next two subsections.

\subsubsection{Hybrid approach} \label{Hybrid quantization approach}

In the case of the hybrid approach to the quantization of primordial perturbations within LQC, the background-dependent masses of the tensor and scalar perturbations, $ s^{(t)}$ and $ s^{(s)}$ respectively, are given by \cite{hybrid_rev,NBMmass}
\begin{eqnarray} \label{eq_hyb_mass} 
    s^{(t)} =  - \frac{4\pi }{3} a^2 (\rho-3P) , \qquad  \quad s^{(s)} = s^{(t)} + \mathcal{U}, \qquad    \quad   \mathcal{U}= a^2  \left[ V_{,\phi \phi} + 48\pi V + 6\frac{a' \phi '}{a^{3}\rho}V_{,\phi} - \frac{48\pi}{\rho}V^{2} \right].
\end{eqnarray}

Two important points should be noted about these masses. First, the scalar and tensor masses coincide with their counterparts in GR when the quantum corrections to the background are switched off. The effects of quantum geometry are therefore only significant in the vicinity of the bounce. Second, the contributions of the inflaton potential become relevant only after the onset of inflation, which allows us to neglect them during the preceding epochs. Consequently, the tensor and scalar masses can be considered effectively identical in the pre-inflationary epochs, for the backgrounds in which we are interested \cite{NM,MVY}. In the following, the treatment of tensor and scalar perturbations will be the same until the onset of inflation, where the two masses acquire distinct behaviours.

We then proceed to solve the mode equations during the three epochs described above. In the quantum bounce epoch, the background-dependent mass leads to a differential equation that cannot be solved analytically. To obtain an analytic solution, we adopt a Pöschl-Teller approximation of the form \cite{NM,waco,waco2}
\begin{equation} \label{PT_hybrid}
  s_{\text{PT}}=\frac{U_0}{\cosh^2(\alpha \eta)}.
\end{equation}
To determine $U_0$ and $\alpha$, we impose that the Pöschl-Teller effective mass coincides with the background-dependent mass at the bounce, obtaining $U_0 = 8\pi\rho_c/3$, and that it matches the mass in GR at the end of the bounce period, getting $\alpha={\rm arccosh}(a_0^2)/\eta_0$ \cite{NM}. This approximation ensures a relative error below $20\%$ throughout the entire bounce epoch when the end of this epoch is chosen to optimize the approximation across both the bounce era and the subsequent classical kinetic period \footnote{\label{PTfoot}The Pöschl–Teller approximation has also been studied in other contexts, such as de Sitter spacetimes \cite{Maciej}, and with alternative choices of vacuum states \cite{waco,waco2}, in both cases yielding results compatible with those presented here. We plan to study other approximations during the bounce epoch and their influence in the results in future works.}. Numerical analyses have shown (see Ref. \cite{NM}) that this procedure suggests a(n approximate) proper time $t_0=0.4$ in Planck units. The solution of the perturbative modes can be found in Appendix \ref{Appendix_Hybrid}.

On the other hand, it is still necessary to impose initial conditions on the modes to obtain their evolution. However, fixing such initial conditions is equivalent to specifying a vacuum state. As mentioned in the Introduction, in models with a pre-inflationary epoch, the choice of a vacuum becomes a nontrivial task, because in generic non-stationary curved spacetimes, such as those arising in pre-inflationary scenarios, there is no preferred state based only on arguments of symmetry invariance. In this work, we adopt the NO-AHD proposal (as commented in the Introduction), which selects a vacuum well adapted to the background, guarantees a good ultraviolet behavior, and suppresses spurious oscillations in the PPS \cite{NMT}.  Using the Pöschl-Teller approximation and imposing the asymptotic diagonalization condition (inherent to the NO-AHD proposal) during the bounce epoch, it is possible to determine the vacuum state, as shown in Ref. \cite{NM} and explained in Appendix \ref{Appendix_Hybrid}. According to our previous comments, the vacuum state selected in this way is optimally adapted to the background dynamics during the bounce period. Since we can calculate it analytically throughout this period \cite{NM}, we can find its value and that of its derivative at any time in this interval. To simplify the analysis and avoid redundant calculations, we compute these values at the end of the bounce epoch, thus getting initial conditions at that instant to fix the rest of the vacuum evolution.

The second period, the purely kinetic period, is well described by the background-dependent mass of GR in absence of an inflaton potential, a case for which the exact solutions for the modes are known. In order to fix the specific solution that corresponds in this period to the NO-AHD vacuum that we have previously determined around the bounce, we impose continuity of the mode solutions up to their first derivatives at the conformal time $\eta_0$ (for more details see Appendix \ref{Appendix_Hybrid} and Refs. \cite{NM,NMY}). The third and final period that we consider is a slow-roll inflationary stage. During inflation, scalar and tensor masses differ in the value of the order $\nu$ of the Hankel functions that provide the mode solutions. For the tensor mass, we have $\nu_{(t)}= \sqrt{3 \epsilon_V +9/4}$, whereas in the scalar case we obtain $\nu_{(s)} = \sqrt{ 9\epsilon_V - 3 \delta_V+9/4}$. These mode solutions are given in Appendix \ref{Appendix_Hybrid}, as well as the specific solution corresponding to our choice of vacuum, which is again calculated by imposing continuity in the modes up to the first derivative at the conformal time $\eta_i$ \cite{NMY,MVY}. Furthermore, in Appendix \ref{Appendix_Transformation}, we offer an alternative but equivalent method for calculating the overall mode solutions for the various periods using transformation matrices, which may be useful for calculations if one wants to vary the number of periods or the properties of individual periods.

\subsubsection{Dressed metric approach}

In the case of the dressed metric approach to the quantization of tensor and scalar perturbations within LQC, the background-dependent masses for the tensor and scalar perturbations, $\bar{s}^{(t)}$ and $\bar{s}^{(s)}$, are given by \cite{NBMmass} 
\begin{eqnarray} \label{eq_hyb_mass}
    \bar{s}^{(t)} = - \frac{4\pi}{3}a^{2}\rho \left(1 + 2\frac{\rho}{\rho_{c}} \right) + 4\pi a^{2}P \left(1 - 2\frac{\rho}{\rho_{c}}   \right) =-\frac{a^{\prime\prime}}{a} , \qquad  \quad \bar{s}^{(s)} = \bar{s}^{(t)} + \mathcal{V},    
\end{eqnarray}
where we have used the second identity in Eq. \eqref{eq_Friedman_LQC} to rewrite $\bar{s}^{(t)} $ in terms only of the scale factor, and 
\begin{eqnarray}
    \mathcal{V}= a^2 \left[V_{,\phi\phi} +48\pi  V - \frac{48\pi }{\rho} V^2 - \sigma_a \sqrt{\frac{96\pi }{\rho}} \frac{|\phi'|V_{,\phi}}{a} \right]. 
\end{eqnarray}
Here, $\sigma_a$ is the sign of the product between $\phi'$ and the canonical momentum of the scale factor. The expressions for the tensor and scalar masses differ from their counterparts in the hybrid case. This difference is significant around the bounce, where quantum effects are most relevant. However, it disappears when the dynamics enter the classical kinetic regime. Therefore, the masses of the scalar and tensor perturbations can be treated in the same way and with the same approximations as in the hybrid analysis after the bounce period, leading to the same general mode solutions, but differing in the specific solution that corresponds to the NO-AHD vacuum determined around the bounce \cite{NM,AMV}. In this subsection, we hence address only the differences during the bounce period.

In this period, the dressed metric mass takes negative values very near the bounce before becoming positive. To accommodate this behavior, we again employ the Pöschl–Teller approximation used in the hybrid case, but with modified parameters and an additional constant term. This constant term allows the approximation to account for the fact that, in the dressed metric approach, the background-dependent mass changes sign during the phase where quantum effects are still significant. The Pöschl–Teller approximation reads \cite{AMV}
\begin{equation}\label{PT_dress}
        \bar{s}_{\text{PT}}(\eta)= \frac{\bar{U}_0-\bar{v}_0}{\cosh^2(\bar{\alpha} \eta)} + \bar{v}_0.
\end{equation}
The values of $\bar{U}_0$, $\bar{v}_0$, and $\bar{\alpha}$ are determined by imposing three conditions. The first two conditions require that the Pöschl–Teller approximation coincides with the exact dressed metric mass at the bounce and at the instant when this mass vanishes, which in proper time corresponds to $t_{\star}  = \sqrt{1/(8\pi \rho_c)} $. The third condition requires that the Pöschl–Teller approximation matches the GR mass at the start of the pure kinetic regime, i.e., at proper time $t_0$. 

The first condition directly gives $\bar{U}_0 = - 8\pi \rho_c$. As for the other two conditions, let us first note that the contribution of the inflaton potential to the background-dependent mass can be ignored during the bounce phase for the solutions of phenomenological interest that we are considering, so that we can use the expression $-a^{\prime\prime}/a$ for both the tensor and the scalar masses. This, together with the form of the scale factor given in Eq. \eqref{scale_factor_LQC} and the value of $\bar{U}_0$, allows us to conclude that $\bar{v}_0$ and $\bar{\alpha}$ must solve the following pair of equations:
\begin{eqnarray}
\label{dressedrelations}
\bar{v}_0 \left[\cosh^2(\bar{\alpha}\eta_{\star})-1\right] &=& 8\pi \rho_c , \\
\bar{v}_0 \left[\cosh^2(\bar{\alpha}\eta_0)-1\right] &=& 8\pi \rho_c \left [1+\frac{1}{3a_0^{4}}\cosh^2(\bar{\alpha}\eta_0)\right].
\end{eqnarray}
Here, $\eta_{\star}$ is the conformal time corresponding to $t_{\star}$ according to relation \eqref{confytime}. Hence, $\eta_{\star}$ is totally determined by the value of $\rho_c$. The relation \eqref{confytime} also allows us to express $\eta_0$ in terms of $a_0$ and the critical density, since Eq. \eqref{scale_factor_LQC} implies that $t_0=\sqrt{(a_0^6-1)/(24\pi\rho_c)}$. To ensure a relative error below $20\%$ between the Pöschl–Teller approximation and the exact dressed metric mass, we set the end of the bounce period at $t_0 = 0.825$ in Planck units in the dressed metric case and compute with this value the remaining parameters. We note that this proper time is larger than the end of the bounce period in the hybrid approach ($t_0 = 0.4$). This difference arises because the corresponding background-dependent masses are different during the bounce period. In the dressed metric case, this mass is negative at the bounce (instead of positive, as it happens in the hybrid approach), so that the modes that are sensitive to quantum effects remain affected for a longer interval, until the mass approaches its kinetic counterpart, which is strictly positive. Consequently, the two approaches lead to different values of $t_0$. 

The general mode solution is the same as in the hybrid case but using the transformation $k \to \bar{k} = \sqrt{k^2 + \bar{v}_0}$ inside the $k$-mode solution. This correspondence applies as well for the initial conditions that select the NO-AHD vacuum in the bounce period \cite{AMV}.

During the kinetic period and the inflationary stage, the scalar and tensor masses become the same as in the hybrid approach, as mentioned before. The general mode solutions are then identical in both cases in those periods; nevertheless, since the NO-AHD vacuum solution changes in the bounce period
(and also the duration of this period changes), 
the specific mode solution corresponding to it in the rest of periods differs, leading to different results when the modes freeze during inflation \cite{MVY,MVY2}. For more details on the expressions of the vacuum mode solution, see Appendix \ref{Appendix_Dressed}.

\section{Primordial power spectrum  in loop quantum cosmology}

\subsection{Definition of primordial power spectrum}

In principle, the generation of the PPS is complete at the end of inflation. However, after a mode exits the Hubble horizon, it looses causal contact with the expanding observable region of the universe and its evolution becomes frozen, in the sense that the amplitude of the mode becomes constant. Therefore, for all practical purposes, the PPS can be evaluated at the time $\eta_f$ when all the modes in a given observable window of the CMB have frozen. In fact, numerical computations indicate that, around the time the last detectable modes freeze and for the kind of background solutions that we are considering, the slow-roll parameters take small, slowly varying values. This supports the assumption that the freezing time $\eta_f$ occurs in the slow-roll regime. The PPS for the scalar and tensor perturbations are then given by \cite{Langlois, Baumann}
\begin{eqnarray}\label{eq_def_PPS}
        \mathcal{P}_\mathcal{R} = \frac{k^{3}}{2\pi^2} \frac{| \mu_k^{(s)}
        (\eta_{f})|^2}{z(\eta_{f})^2} \qquad \text{and} \qquad         \mathcal{P}_\mathcal{T} = \frac{32k^{3}}{\pi} \frac{|\mu_k^{(t)}
        (\eta_{f})|^2}{a(\eta_{f})^2},
\end{eqnarray}
respectively, where $z = a^2\phi'/ a'$ which in the slow-roll regime becomes $z^2 = a^2 \varepsilon_V/ (4\pi)$. 

We can also use the slow-roll approximation to rewrite in a much simpler form the mode solution for the NO-AHD state that we have determined around the bounce (given in Eq. \eqref{mu_k_SR_hybrid}). This is possible because the arguments of the Hankel functions that appear in that mode solution are much smaller than one, at least for wave numbers in a sufficiently large interval which includes the observable window. Employing the asymptotic behavior of the Hankel functions \cite{Abra}, we then obtain the approximated expression 
\begin{eqnarray}
        \left|\mu_k^{SR}\right|^2 = \frac{1}{4\pi} (\eta_{e}-\eta_{f}) |\Gamma(\nu_{(t,s)})|^2 \left[\frac{k(\eta_{e}-\eta_{f})}{2}\right]^{-2\nu_{(t,s)}} \left|A_k^{(t,s)} - B_k^{(t,s)}\right|^2,
\end{eqnarray}
where $\Gamma$ is the gamma function, we have used the pair of labels $(t,s)$ to denote simultaneously the scalar and tensor cases, the label $SR$ stands for slow roll, and we recall that $\eta_e$ is the conformal time at the end of inflation. In addition, $A_k$ and $B_k$ (ignoring the $(t,s)$ labels) are the constants in the linear combination of Hankel functions that determine the vacuum solution for the modes during the slow-roll phase. 

Introducing the pivot scale $k_*$, defining $k_f = (\eta_e - \eta_f)^{-1}$ as the last mode in our observable window to cross the horizon, and employing that Eq. \eqref{scale_factor_SR} implies that $k_f = a(\eta_f)H(f )$ at dominant order in the slow-roll approximation, the PPS becomes
\begin{eqnarray}
    \mathcal{P}_\mathcal{R} = A_{s} \left(\frac{k}{k_*
    }\right)^{n_s-1}   \left|A_k^{(s)}-B_k^{(s)}\right|^2, \qquad\mathcal{P}_\mathcal{T} = A_{t} \left(\frac{k}{k_*
    }\right)^{n_t}\left|A_k^{(t)}-B_k^{(t)}\right|^2\ , 
\end{eqnarray}
where as usual we have defined the scalar and tensor spectral indices as $n_s-1=3-2\nu_{(s)}$ and $n_t=3-2\nu_{(t)}$, and the scalar and tensor amplitudes are $A_{s} = \left(k_*/k_f\right)^{n_s-1} H_f^2/ (\pi \epsilon_V) $ and $A_{t}=16 \left(k_*/k_f\right)^{n_t} H_f^2/\pi$ respectively, with $H_f$ being the value of the Hubble parameter when the modes freeze \cite{Langlois, Baumann}. In order to compare our results with those of the Planck collaboration, we adopt the same pivot scale $k_*$ used in the Planck analysis. The pivot scale represents an arbitrary reference scale, chosen to match scales to which a given experiment is most sensitive or to minimize correlations among model parameters in the parameter interference \cite{Planck_2013_params,Planck_2015_params,Planck:2015sxf}.

We emphasize that the dependence of the PPS on the constants $A_k$ and $B_k$ of the modes relates directly to the choice of vacuum state. The existence of a pre-inflationary epoch has a direct consequence on the PPS, since the  Bunch-Davies vacuum is no longer the preferred vacuum state, owing to the evolution of the modes away from a quasi-de Sitter regime during the bounce period and the subsequent kinetic period prior to inflation. Different choices of vacuum state to account for this pre-inflationary physics lead to different spectra in general. 

Furthermore, the above expressions for the PPS typically display fast oscillations in $k$ because, even if the norms of $A_k$ and $B_k$ usually vary slowly, their phases give rise to a rapidly oscillating interference\footnote{Such rapid oscillations contribute to an enhancement of power in the angular power spectrum, thereby preventing the suppression of power at low scales and hence a better fit to the data.}. These fast oscillations can be attributed to the instantaneous transition between the different considered periods and to the fact that our choice of vacuum is only optimally adapted to the bounce era, but not to the rest. However, it is possible to remove the spurious oscillations in $k$ produced by this mismatch in our choice of vacuum: as argued and supported in Ref. \cite{NM}, we can adjust our choice of vacuum using the following Bogoliubov transformation, that leads to a state truly free of unwanted oscillations in the PPS :
 \begin{eqnarray}
A_k^{(t,s)} \to \tilde{A}_k^{(t,s)} = \left|A_k^{(t,s)}\right|, \quad \quad B_k^{(t,s)} \to \tilde{B}_k^{(t,s)} = \left|B_k^{(t,s)}\right|.
\end{eqnarray}
This Bogoliubov transformation connects the two states and eliminates the undesired oscillations in the spectrum of the originally considered vacuum. From now on, we assume that this transformation has been implemented and ignore the tilde in our notation for the mode constants. The corresponding scalar and tensor PPS are simply
\begin{eqnarray}
\mathcal{P}_\mathcal{R} =  \left(\left|A_k^{(s)}\right|-\left|B_k^{(s)}\right|\right)^2 \; \mathcal{P}_\mathcal{R}^{\text{$\Lambda$CDM}} , \qquad    \mathcal{P}_\mathcal{T} = \left(\left|A_k^{(t)}\right|-\left|B_k^{(t)}\right|\right)^2 \; \mathcal{P}_\mathcal{T}^{\text{$\Lambda$CDM}}.
\end{eqnarray}
where $ \mathcal{P}_\mathcal{R}^{\text{$\Lambda$CDM}}  =  A_s \left(k/k_*\right)^{n_s-1}$ and $ \mathcal{P}_\mathcal{T}^{\Lambda\rm{CDM}}  =  A_t \left(k/k_*\right)^{n_t}$ are the standard expressions of the PPS in the $\Lambda$CDM model for slow-roll inflation. The prefactors multiplying these standard spectra in our equations can clearly be interpreted as due to the Bogoliubov transformation between the Bunch-Davies state and our choice of vacuum at the time when all the considered modes have frozen during slow-roll inflation.

\subsection{Parametrization of the pre-inflationary factor}

We now want to reach a concise and manageable parametrization of the factor that corrects the $\Lambda$CDM spectra. With this aim, we will justify some approximations that allow us to simplify the expressions for $|A_k|$ and $|B_k|$ in the hybrid and dressed metric approaches (these expressions are provided in the Appendix: see Appendix \ref{Appendix_Hybrid} and Appendix \ref{Appendix_Dressed}). We wish to reduce as much as possible the number of parameters on which $|A_k|$ and $|B_k|$ depend, to better isolate the
relevant characteristics of the pre-inflationary period for the CMB. For this purpose, an interesting possibility is to further use the asymptotic behaviour of the Hankel functions (which appear in $|A_k|$ and $|B_k|$) if the scales involved in the model investigated in this work justify the asymptotics. For this, we must analyze the arguments of those functions and determine under which conditions the asymptotic behaviour is approximately valid.

We first notice that the interval of wave numbers of interest should contain the suppression scale in order to include relevant quantum corrections while allowing for modes compatible with the extremely good fit of the observations that the standard $\Lambda$CDM model provides. This suppression scale approximately corresponds to $k_0 = a_0H_0$. We must then compare $k_0$ with the other characteristic scales of the model. These are the pure kinetic scale $k_i=a_0^3H_0/a_i^2=a_iH_i$, given by the product of the scale factor $a_i$ and the Hubble parameter $H_i$ at the end of the kinetic period (which coincides with the onset of inflation), and the inflationary scale\footnote{We note that this definition of the inflationary scale $k_e$ does not coincide with the scale determined by the product $a_eH_e$. Nonetheless, we use the symbol $k_e$ for such scale to avoid overcomplicating the notation.}  $k_e=1/(\eta_e-\eta_i)$ (see Appendix \ref{Appendix_Hybrid}). Actually, at dominant order in the slow-roll approximation we have that $k_e = 1/(\eta_e - \eta_i) \approx a_i H_i $, so that both scales can be approximately identified. 

Comparing then $k_i$ with $k_0$ and using the expression of the Hubble parameter during the kinetic period to compute $H_i$, we obtain 
\begin{equation}
    \frac{k_i}{k_0} = \frac{a_i H_i}{a_0 H_0} = \left( \frac{H_i}{H_0} \right)^{2/3}.
\end{equation}
From the CMB observations, it is estimated that the Hubble parameter at the end of the kinetic period (that is, at the onset of slow roll within our approximated description) is $H_i \approx 0.6 \times 10^{-6} H_0$ \cite{Liu}, which implies $k_0 \approx 10^{4}\, k_i$. Consequently, for wave numbers not very much smaller than the suppression scale, we have $k/k_i \gg 1$, and henceforth also $k/k_e\gg 1$. We remark that the observable window of modes cannot correspond to much smaller values of $k$ without a conflict, because this would imply the prediction of significant quantum corrections for most of the modes detected in measurements of the CMB. On the other hand, an observable window with values of $k$ much larger than the supression scale $k_0$, although observationally allowed in principle, would result in a nearly scale-invariant PPS without traces of LQC effects, essentially reproducing the same predictions as slow-roll inflation.

We may then resort to the asymptotic behaviour of the Hankel functions for large arguments in $k/k_i$ or $k/k_e$, and subsequently take the limit in which these two ratios tend similarly to infinity. In this way, we find the following expressions for the norms of $A_k^{(t,s)}$ and $B_k^{(t,s)}$ in the hybrid case:
\begin{eqnarray}\label{eq_Ak_hyb_param}
      \nonumber \left | A_k^{(t,s)} \right | &=&  \frac{1}{4} \sqrt{ {\frac{\pi k}{k_0}}}\, \Biggl |  \frac{\alpha b^k_M}{2a_0^4k}  H_0^{(1)}\left[\frac{k}{2k_0} \right]  \,{}_{2}F_1 \left( b_1^k+1,b_2^k+1,b_3^k+1 ;\frac{1+\sqrt{1-a_0^{-4}}}{2}\right) 
     \\
     &-& \left \{ \left (\frac{k_0}{k} - i\sqrt{1-a_0^{-4}}\right) H_0^{(1)}\left[\frac{k}{2k_0} \right]  -H_1^{(1)}\left[\frac{k}{2k_0} \right] \right \}  {}_{2}F_1\left(b_1^k,b_2^k;b_3^k;\frac{1+\sqrt{1-a_0^{-4}}}{2}\right)  \Biggl | ,
\end{eqnarray}
\begin{eqnarray}\label{eq_Bk_hyb_param}
        \left | B_k^{(t,s)} \right | &=& \left | A_k^{(t,s)}\right |_{(1) \leftrightarrow (2)}  \,,
\end{eqnarray} 
where $H_{\nu}^{(n)}$ is the Hankel function of $\nu$th-order and $n$th-kind \cite{Abra}, and in the last formula we have used a compact notation to indicate the interchange of Hankel functions of the first and second kind in the corresponding expression of $|A_k|$ \footnote{The asymptotic expansions of the Hankel functions that we have used are valid within the observable window considered in this work. In cases where one might be interested in low values of $k$ for which this asymptotic behavior does not hold, the exact expressions given in Appendix \ref{Appendix_A} remain valid and can be used.}.

From these equations, we note that the final expression depends directly on two parameters: the value of the scale factor at the end of the bounce period, $a_0$, and the suppression scale, $k_0 = a_0 H_0$. The constants $b_j^k$ ($j=1, 2, 3$) of the hypergeometric functions are given in Appendix \ref{Appendix_Hybrid}, and depend directly on $a_0$ and $k_0$, as it is easy to check from Eq. \eqref{eq_b_i}.

Furthermore, the suppression scale $k_0$ is related to $a_0$ by the value of the energy density at the bounce, the critical density $\rho_c$. Indeed, taking into account that the quantum bounce is also kinetically dominated in our background solutions of interest, we get $k_0 = \sqrt{8\pi\rho_c/(3a_0^{4}) }$. 

By a similar procedure, we can obtain the expression of the norm of the mode coefficients in the case of the dressed metric approach (that we distinguish with an overbar),
\begin{eqnarray}\label{eq_Ak_dress_param}
      \nonumber \left | \bar{A}_k^{(t,s)} \right | &=&  \frac{1}{4} \sqrt{\frac{\pi k^2}{\bar{k} k_0}}\, \Biggl |  \frac{2\bar{\alpha} \bar{b}^{\bar{k}}_M}{k} \bar{x}_0(1-\bar{x}_0)  H_0^{(1)}\left [ \frac{k}{2k_0} \right ] \;{}_{2}F_{1}\left( \bar{b}_{1}^{\bar{k}}+1, \bar{b}_{2}^{\bar{k}}+1, \bar{b}_{3}^{\bar{k}}+1; \bar{x}_0 \right)\\
      &-&  \left \{ \left (i \frac{\bar{k}}{k}(1-2\bar{x}_0)+  \frac{k_0}{k}\right) {H_0^{(1)}\left [\frac{k}{2k_0}\right ]} - {H_1^{(1)}\left [ \frac{k}{2k_0}\right]} \right \}  \; {}_{2}F_{1}\left( \bar{b}_{1}^{\bar{k}}, \bar{b}_{2}^{\bar{k}}, \bar{b}_{3}^{\bar{k}}; \bar{x}_0 \right)\Biggl | ,
\end{eqnarray}
\begin{eqnarray}\label{eq_Bk_dress_param}
        \left | \bar{B}_k^{(t,s)} \right | &=& \left | \bar{A}_k^{(t,s)} \right |_{(1) \leftrightarrow (2)} \,.
\end{eqnarray}

In these formulas, $\bar{\alpha}$ and $\bar{v}_0$ are functions of $a_0$ and $k_0$ (or equivalently, of $a_0$ and the critical density), as they must satisfy relations \eqref{dressedrelations}. The quantities $\bar{b}^{\bar{k}}_j$ ($j=1, 2, 3$) and $\bar{x}_0$, apart from $k$, depend only $\bar{\alpha}$, $\bar{v}_0$ and the critical density, and hence also only on the two independent parameters $a_0$ and $k_0$. They are given in Appendix \ref{Appendix_Dressed}.

At this stage, it should be noted that if, instead of setting the scale factor at the bounce to unity, we adopt the convention of making the current scale factor equal to one, all scales must be changed accordingly using the total number of e-folds accumulated from the bounce to the present day, as we will discuss in the next subsection.

\begin{figure}[h!]
            \begin{minipage}[c]{.48\linewidth}
            \centering
                \includegraphics[width=10cm]{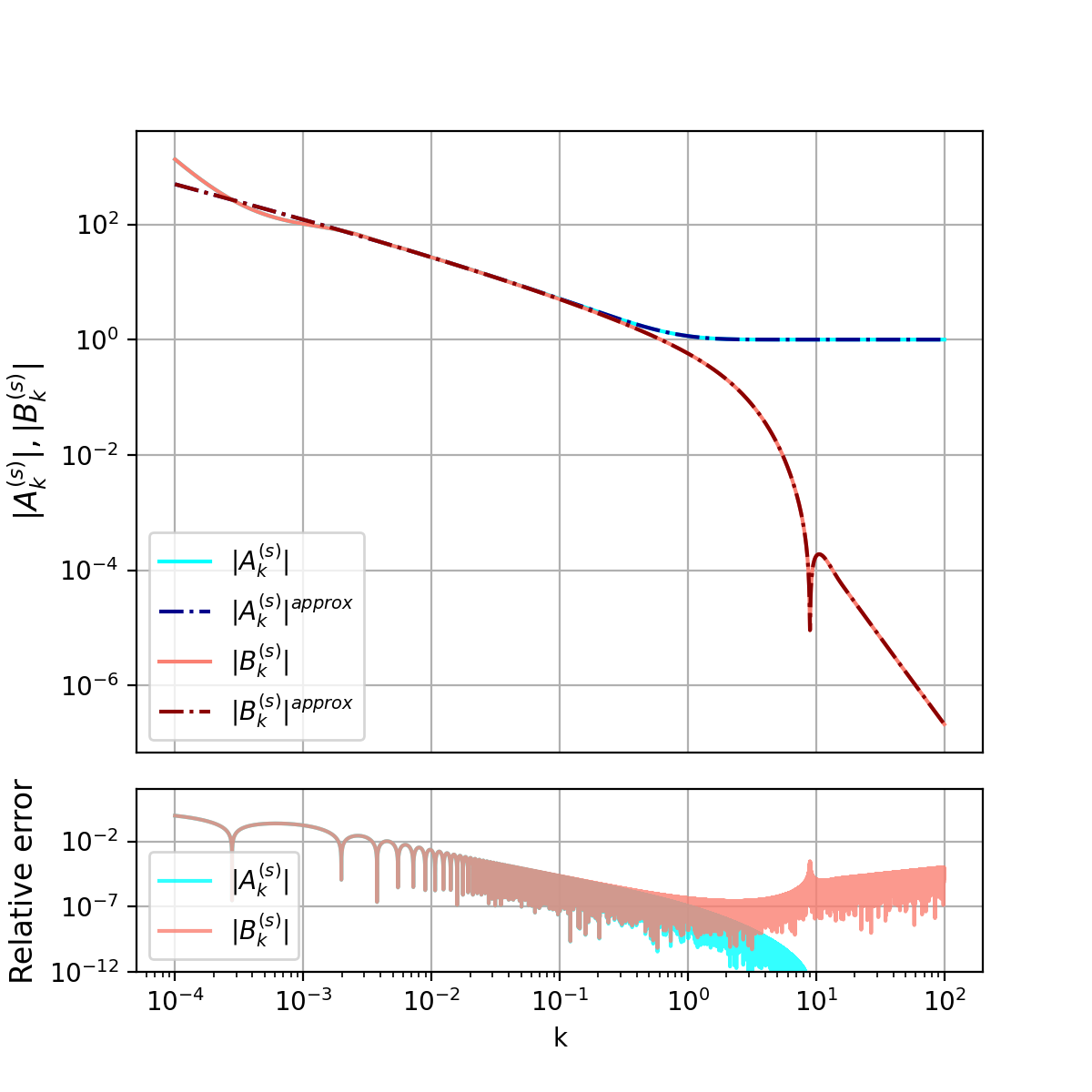}
                \caption{Norms of the constants $A_k^{(s)}$ and $B_k^{(s)}$ for the scalar perturbations in the hybrid approach. The solid and the dashed lines correspond respectively to expressions \eqref{eq_Ak_hyb} and \eqref{eq_Bk_hyb}, where no approximation based on relations between scales is used, and to expressions \eqref{eq_Ak_hyb_param} and \eqref{eq_Bk_hyb_param}, where such approximation is employed. We have taken $\gamma=0.2375$ and $\phi_0=0.97$ for the Immirzi parameter and the value of the inflaton at the bounce, respectively, and considered a quadratic inflaton potential with mass $m=1.2 \times 10^{-6}$.}
                \label{fig_Comparison_Ak_Bk_Hyb}
            \end{minipage}
            \hfill
            \begin{minipage}[c]{.48\linewidth}
            \centering
                \includegraphics[width=10cm]{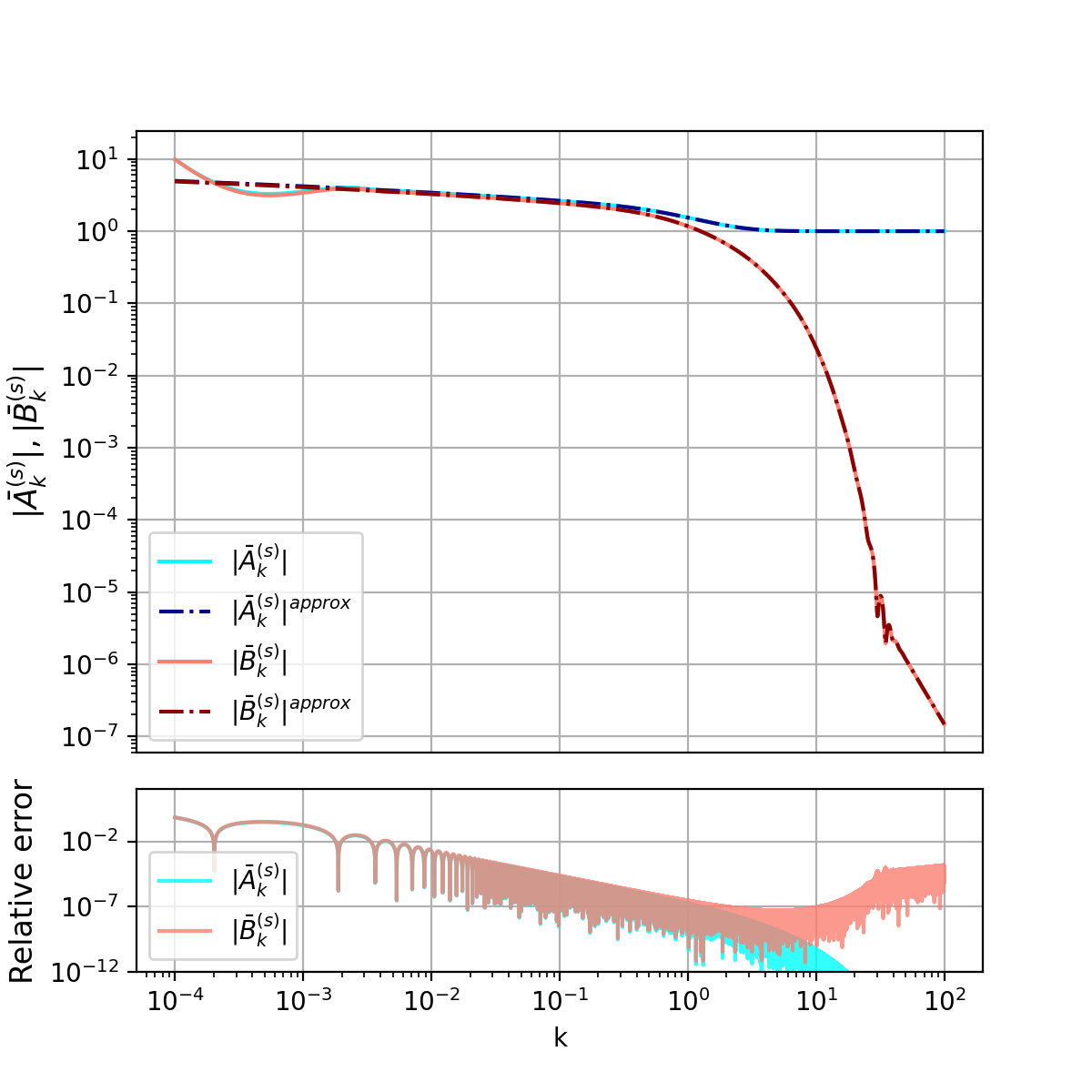}
                \caption{Norms of the constants $\bar{A}_k^{(s)}$ and $\bar{B}_k^{(s)}$ for the scalar perturbations in the dressed metric approach. The solid and the dashed lines correspond respectively to
expressions \eqref{eq_Ak_dress} and \eqref{eq_Bk_dress}, where no approximation based on relations between scales is used, and to expressions \eqref{eq_Ak_dress_param} and \eqref{eq_Bk_dress_param}, where such approximation is employed. We have taken $\gamma=0.2375$ and $\phi_0=0.97$ for the Immirzi parameter and the value of the inflaton at the bounce, respectively, and considered a quadratic inflaton potential with mass $m=1.2 \times 10^{-6}$.}
                \label{fig_Comparison_Ak_Bk_DM}
                \end{minipage}
        \end{figure}

For a background with inflaton at the bounce equal to $\phi_0=0.97$ (a value used in previous works\footnote{On the other hand, numerical results indicate that the sensitivity to the choice of $\phi_0$ is not very high.}, thus enabling direct comparisons \cite{hybridCMB,NM,MVY,MVY2}) and subject to a quadratic potential, the norms of $A_k$ and $B_k$ for scalar perturbations with our parametrization are represented in Fig. \ref{fig_Comparison_Ak_Bk_Hyb} and Fig. \ref{fig_Comparison_Ak_Bk_DM}, respectively for the hybrid approach and the dressed metric approach. In both cases, the result adjusts very well to the corresponding formulas without approximations based on the relative values of the involved scales. The relative errors, for both approaches, are below $1\%$ for an ample interval of wave numbers, including a large region around the scale suppression $k_0$. The same conclusion is also reached in the case of the tensor perturbations.    

\subsection{Parametrized Primordial Power Spectrum}
\label{sec:PPPS}

We now have all the ingredients needed to compute and compare the PPS obtained with our parametrization, on the one hand, and with the general formulas without approximations based on the relative magnitude of the involved scales, on the other hand. Before doing that, let us first comment on the window of observable modes and its translation to the model investigated in this work. The observable modes approximately correspond to the interval $2 \times 10^{-4}  {\rm Mpc}^{-1} \leq k / a_{\rm today} \leq 6 \times 10^{-1}  {\rm Mpc}^{-1}$ in inverse megaparsecs \cite{Planck_Constraints}, using the standard convention of setting the value of the scale factor today, $a_{\rm today}$, equal to 
1 in order to fix scales. By contrast, recall that in our discussion we have set to 1 the value of the scale factor at the bounce, to simplify our calculations. To relate both conventions, we must use that $a_{\rm today} =a_B e^{\Delta n} $, where $\Delta n$ is the number of e-folds during the entire cosmic evolution from the bounce until today. Using this relation and converting inverse megaparsecs to inverse Planck units, we can say that, roughly speaking, the observable window becomes $[1, 3 \times 10^3] e^{\Delta n - 140}$ in Planck units. Moreover, numerical studies carried out in LQC \cite{AG1, Morris} lead to a total number of e-folds in the range $130 \leq \Delta n \leq 143$ for our background solutions of phenomenological interest and, together with observational data regarding the cosmological evolution after inflation, indicate a special preference for a total number of e-folds between 135 and 140, number for which LQC effects have been shown to be able to alleviate CMB tensions \cite{AG1, Morris}. Motivated by these arguments and for the sake of concreteness, we will pay special attention to wave numbers $k \in \left[10^{-4}, 10^2\right]$, which are potentially observable in the considered scenarios with short-lived inflation. It is important to stress that the total number of e-folds determines whether the suppression scale $k_0 \sim 1$ in Planck units lies within the observable window or not.

\begin{figure}[h!]
            \begin{minipage}[c]{.48\linewidth}
            \centering
                \includegraphics[width=10cm]{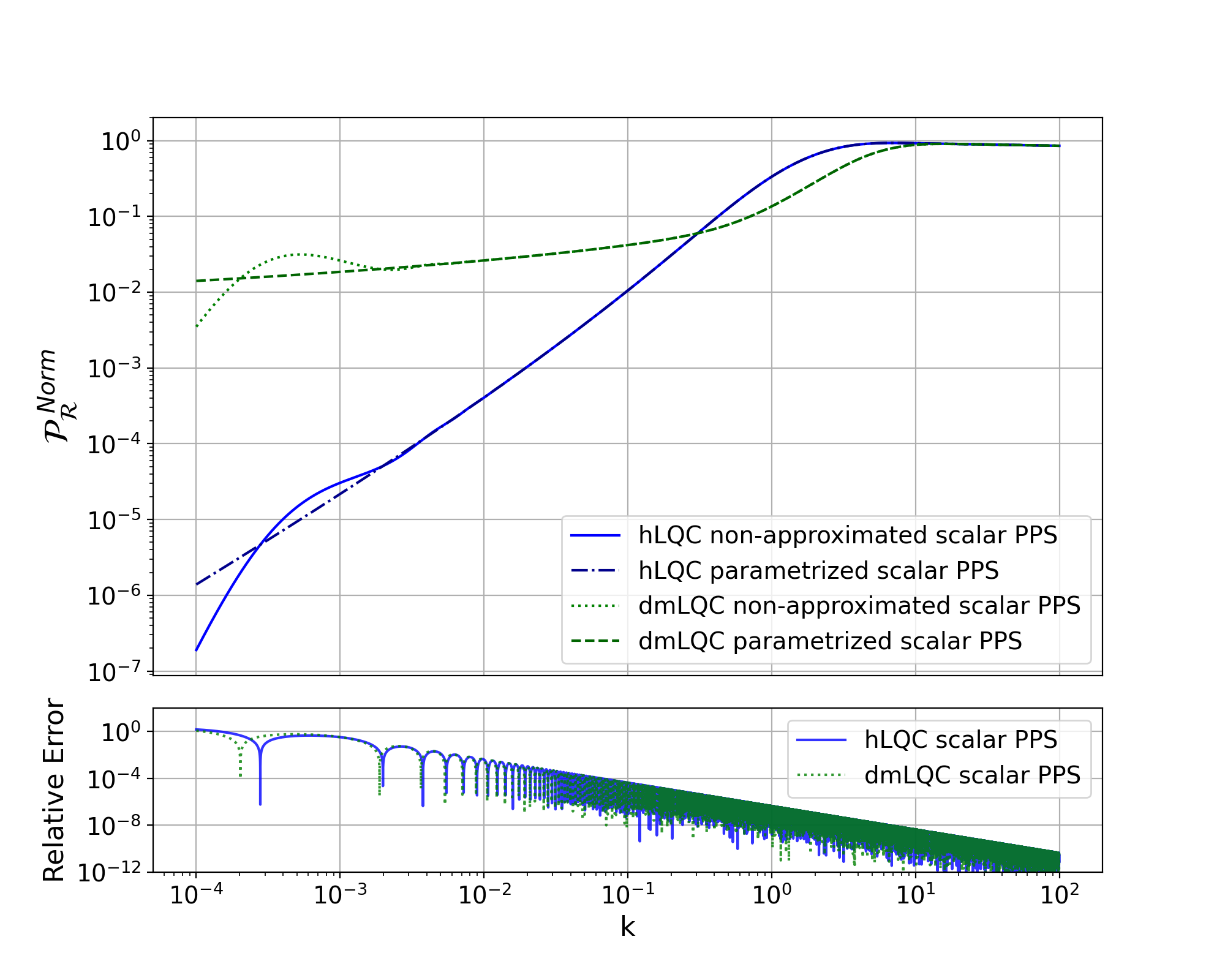}
                \caption{\textbf{Top}: Normalized PPS for scalar perturbations, $\mathcal{P}_\mathcal{R}^{\text{Norm}}$. We show the non-approximated result in the hybrid approach (solid blue line), the parametrized approximation in the hybrid approach (dot-dashed dark blue line), the non-approximated result in the dressed metric approach (dotted green line), and the parametrized approximation in the dressed metric approach (dashed dark green line). In the legends, hLQC stands for the hybrid approach, and dmLQC for the dressed metric case.  \textbf{Bottom}: Relative error between the formula without approximations based on the relative magnitudes of scales and our parametrization in the hybrid approach (solid blue line) and the dressed metric approach (dotted green line). We have taken $\gamma=0.2375$ and $\phi_0=0.97$ for the Immirzi parameter and the value of the inflaton at the bounce, respectively, and considered a quadratic inflaton potential with mass $m=1.2 \times 10^{-6}$.}
                \label{fig_PPS_scalar}
            \end{minipage}
            \hfill
            \begin{minipage}[c]{.48\linewidth}
            \centering
                \includegraphics[width=10cm]{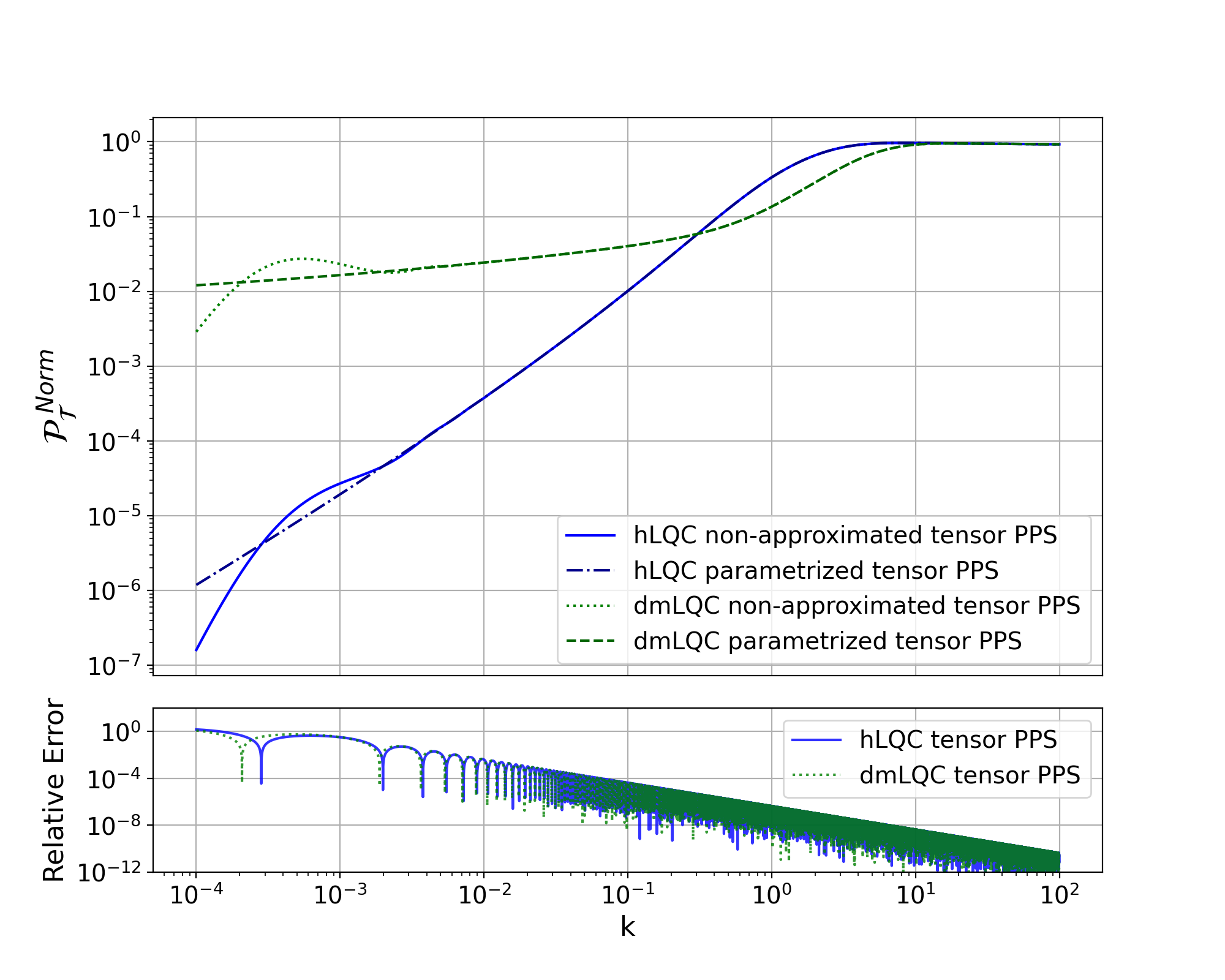}
                \caption{ \textbf{Top}: Normalized PPS for the tensor perturbations, $\mathcal{P}_\mathcal{T}^{\text{Norm}}$. We show the non-approximated result in the hybrid approach (solid blue line), the parametrized approximation in the hybrid approach (dot-dashed dark blue line), the non-approximated result in the dressed metric approach (dotted green line), and the parametrized approximation in the dressed metric approach (dashed dark green line). In the legends, hLQC stands for the hybrid approach, and dmLQC for the dressed metric case. \textbf{Bottom}: Relative error between the formula without approximations based on the relative magnitudes of scales and our parametrization in the hybrid approach (solid blue line) and the dressed metric approach (dotted green line). We have taken $\gamma=0.2375$ and $\phi_0=0.97$ for the Immirzi parameter and the value of the inflaton at the bounce, respectively, and considered a quadratic inflaton potential with mass $m=1.2 \times 10^{-6}$. }
                \label{fig_PPS_tensor}
                \end{minipage}
        \end{figure}

For the interval $k \in \left[10^{-4}, 10^2\right]$ in Planck units, we display in Fig. \ref{fig_PPS_scalar} and Fig. \ref{fig_PPS_tensor} respectively, the normalized PPS for scalar and tensor perturbations
for both the hybrid and the dressed metric approaches, using the general formula without approximations and our corresponding parametrization. To avoid dealing with parameters that are not relevant for our analysis, we define this normalized spectrum as the true PPS divided by $A_s/ k_*^{n_s -1}$ or $A_t/k_*^{n_t}$, respectively in the scalar and tensor cases. We exemplify the discussion with the case of a background determined by the initial condition $\phi_B=0.97$ at the bounce, subject to a quadratic potential. If we compare the scalar PPS between the hybrid and the dressed metric approaches, we note that the suppression scale in the dressed metric case is a bit higher than its counterpart in the hybrid approach, but with smaller suppression rate around it, a difference that becomes more drastic towards the infrared region. These features were already found in other works \cite{MVY,MVY2}, supporting our results. We also 
note that both spectra are almost indistinguishable for wave numbers larger than the suppression scale, towards the ultraviolet region. In that region, the spectra reproduce the results of slow-roll inflation, without significant deviations and with the same scalar and tensor tilts as in standard slow roll \cite{Langlois,Baumann}.

The parametrized expressions of the scalar and tensor normalized PPS, for both approaches, adjust well the normalized PPS obtained without approximations due to the different orders of the involved scales, with a relative error below $1\%$ in the region from the suppression scale $k \sim 1$ towards the ultraviolet. In the case of the infrared region below $k \sim 10^{-3}$, the quality of the approximated parametrized formulas starts to deteriorate. This was expected because such scales become comparable to the kinetic scale $k_i$ and the inflationary scale $k_e$, removed in our approximations. Nevertheless, the fact that this happens in a region where the PPS is deeply suppressed, both with or without our approximations, let us expect that this discrepancy will eventually lead to no relevant physical consequences (as we will later corroborate in the particular example of our calculations of the temperature angular power spectra). 

Taking advantage of the parametrization above, we can also easily consider the implications for the tensor-to-scalar ratio in the hybrid and dressed metric approaches to cosmological perturbations in LQC. The tensor-to-scalar ratio $r$ is defined as the quotient of the tensor and the scalar PPS at the pivot scale $k_{*}$. In the $\Lambda$CDM model with slow-roll inflation, this ratio is simply $r \equiv A_t / A_s$ \cite{Baumann}. In LQC, on the other hand, the ratio $r$ at the pivot scale takes the form
\begin{equation}
    r = \frac{A_t\ \left(\left|A_{k_{*}}^{(t)}\right|-\left|B_{k_{*}}^{(t)}\right|\right)^2}{A_s\ \left(\left|A_{k_{*}}^{(s)}\right|-\left|B_{k_{*}}^{(s)}\right|\right)^2},
\end{equation}
ignoring the overbar in the dressed metric case to use a compact notation. From our previous discussion, we know that the parametrized values of the mode integration constants obtained with our approximations, justified by the orders of the different scales that are relevant in the model, are in fact equal for the scalar and the tensor perturbations in the two considered approaches. This indicates that, in the region where the approximations used in our parametrization are well justified, that is, far from the infrared region, the tensor-to-scalar ratio becomes the same expression as in the standard $\Lambda$CDM model. 

Let us however note that if our LQC model yielded the true PPS of the Universe independently of our approximations, and if the CMB spectra resulting from those PPS were analyzed with the standard $\Lambda$CDM model beyond the region of validity of those approximations, then observable constraints on the amplitudes $A_s$ and $A_t$ might be biased. And that bias might be different for each amplitude. In that case, constraints on $r$ itself could be biased. A more thorough investigation of this is left to follow-up work.

\section{Angular power spectrum}

To confront the hybrid and dressed metric predictions with CMB observations, we compare the TT (temperature) angular spectrum computed for both approaches with data from the Planck mission. For this purpose, we use the Cosmic Linear Anisotropy Solving System \texttt{CLASS} to obtain the TT angular spectrum \cite{ClassII}. To include the LQC pre-inflationary physics, we employ the parametrized form of the PPS that we have obtained as an external power spectrum in the internal code of \texttt{CLASS}, maintaining the values of the cosmological parameters $\{\Omega_bh^2,\Omega_ch^2, 100\theta_{\text{MC}}, \tau, \ln(10^{10} A_\text{s}), n_{\text{s}}\}$ equal to those of their best fit given by  the TT, TE, EE+lowE+lensing+BAO recombination data \cite{Planck_parameters}. In addition, in our computations we choose to consider a total number of e-folds in the range $\Delta n \in [138, 143]$, motivated by our previous discussion and as a balance to allow for different cosmological evolutions while facing simple manageable computations. 

The interval of wave numbers $k$ that we must analyze in our computation of the TT angular spectrum corresponds to $k / a_{\rm today} \in [10^{-7}, 10^2]  \ \text{Mpc}^{-1}$, which contains the observable window given by the Planck collaboration \cite{Planck_parameters,Planck_Constraints}, but it is larger. This interval is the minimal set required to ensure stable interpolation within the internal routines of \texttt{CLASS}. Although for a low number of total e-folds (among those that we are studying) the left end of this interval is outside the region where we have proven the validity of our approximations in the PPS, we expect this to pose no problem when calculating the angular power spectrum. The reason is that the power is so highly suppressed in such an infrared region that the difference between the descriptions with and without our approximations should not produce any relevant physical effect. Indeed, we have verified numerically that the angular power spectrum obtained from the non-approximated PPS coincides with that obtained using the parametrized PPS for the analyzed case with the lowest number of e-folds, corresponding to the worst possible scenario. Besides, we recall that our approximations are well justified for all wave numbers much larger than the suppression scale, where one actually reaches almost the same results as in the standard $\Lambda$CDM model with slow-roll inflation. Therefore, for simplicity, in what follows we present only the spectra derived from the parametrized PPS, which are much easier to compute. 

\begin{figure}[h]
    \centering
    \includegraphics[width=1\textwidth]{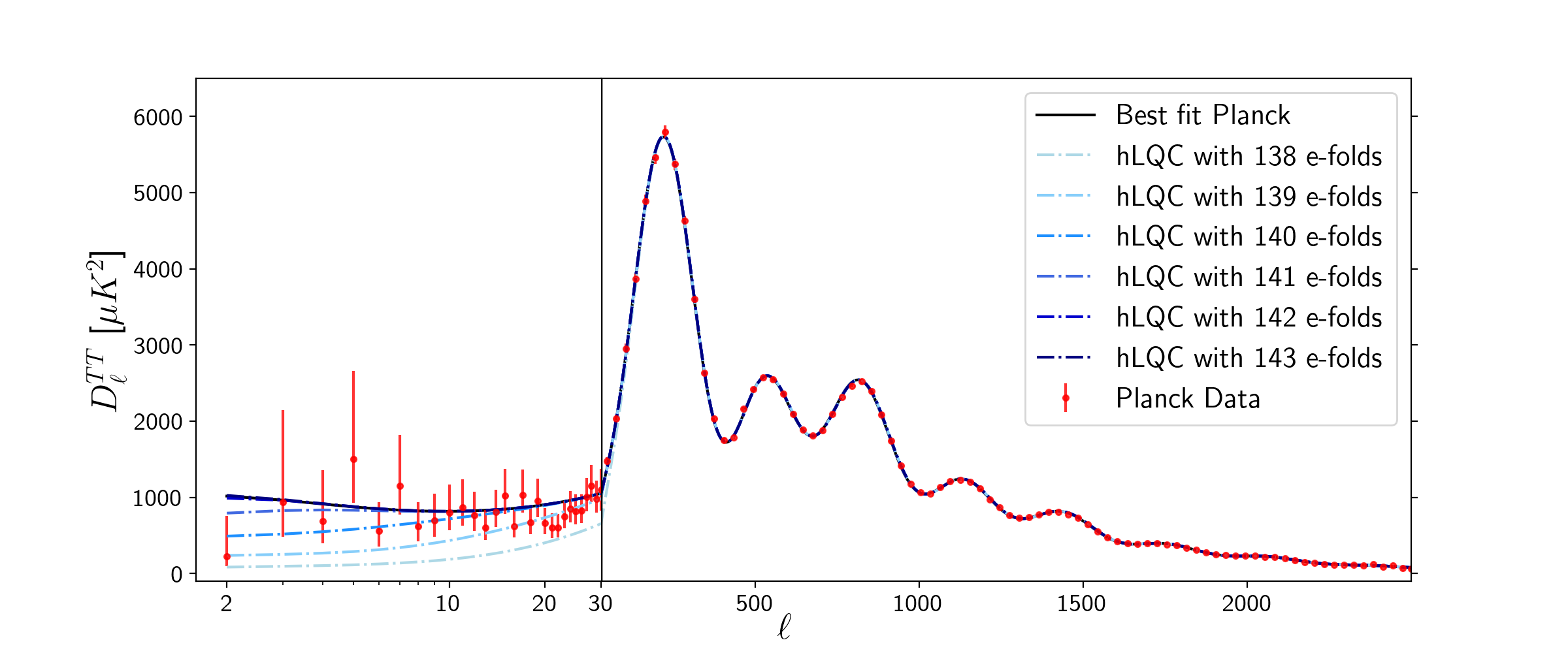} 
    \caption{Temperature angular spectrum, $D_{\ell}^{TT}$. We show the Planck data for $\ell>30$ as the frequency-coadded temperature spectrum computed from the Plik cross-half-mission likelihood, with foreground and other nuisance parameters fixed to a best fit assuming the base-$\Lambda$CDM cosmology obtained by the Planck mission \cite{Planck_parameters,Planck_Constraints}. For low multipoles $2 \leq \ell \leq 30$, we plot the power spectrum estimates from the Commander component-separation algorithm obtained by the Planck collaboration. The solid dark line represents the base-$\Lambda$CDM theoretical spectrum with the best fit to the Planck TT,TE,EE$+$lowE$+$lensing likelihoods of the Planck collaboration. The dot-dashed blue lines represents different angular power spectra obtained from the hybrid approach to LQC (hLQC) for different numbers of total e-folds, computed using the TT,TE,EE$+$lowE$+$lensing$+$BAO estimation of cosmological parameters by the Planck collaboration \cite{Planck_parameters}. We have taken LQC parameters equal to $a_0=1.347$ for the scale factor at the end of the bounce period and $k_0=1.023$ for the suppression scale in Planck units. Note that the vertical scale changes at $\ell$ = 30, where the horizontal axis switches from logarithmic to linear.}
    \label{fig_APS_h}
\end{figure}

\begin{figure}[h]
    \centering
    \includegraphics[width=1\textwidth]{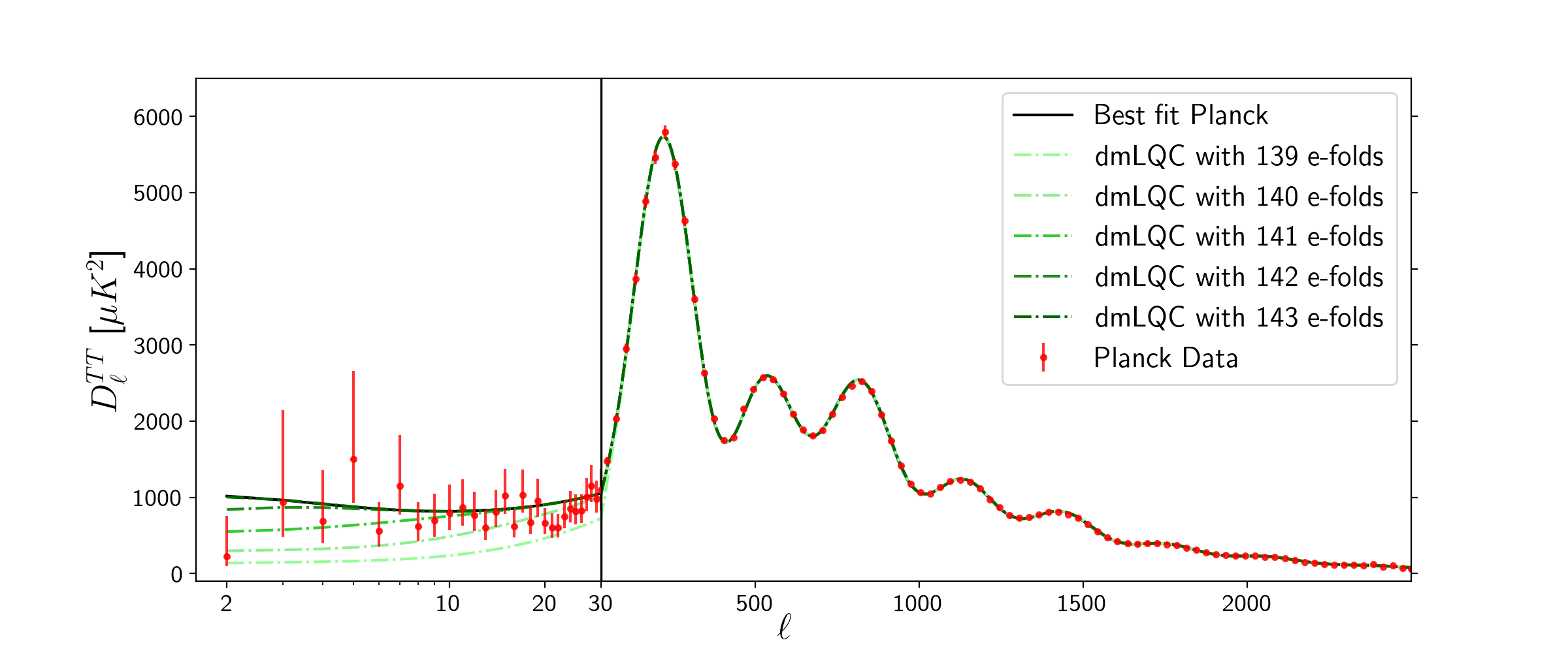} 
    \caption{Temperature angular spectrum, $D_{\ell}^{TT}$. We show the Planck data for $\ell>30$ as the frequency-coadded temperature spectrum computed from the Plik cross-half-mission likelihood, with foreground and other nuisance parameters fixed to a best fit assuming the base-$\Lambda$CDM cosmology obtained by the Planck mission \cite{Planck_parameters,Planck_Constraints}. For low multipoles $2 \leq \ell \leq 30$, we plot the power spectrum estimates from the Commander component-separation algorithm obtained by the Planck collaboration. The solid dark line represents the base-$\Lambda$CDM theoretical spectrum with the best fit to the Planck TT,TE,EE$+$lowE$+$lensing likelihoods of the Planck collaboration. The dot-dashed green lines represents different angular power spectra obtained from the dressed metric approach to LQC (dmLQC) for different numbers of total e-folds, computed using the TT,TE,EE$+$lowE$+$lensing$+$BAO estimation of cosmological parameters by the Planck collaboration \cite{Planck_parameters}. We have taken LQC parameters equal to $a_0 = 1.674$ for the scale factor at the end of the bounce period and $k_0=0.661$ for the suppression scale in Planck units. Note that the vertical scale changes at $\ell$ = 30, where the horizontal axis switches from logarithmic to linear.}
    \label{fig_APS_dm}
\end{figure}

In Figs. \ref{fig_APS_h} and \ref{fig_APS_dm}, we show the TT angular power spectrum for the hybrid and the dressed metric approaches respectively, for different numbers of e-folds in the aforementioned range. On the one hand, we observe that, as the number of e-folds increases, the effects due to LQC corrections disappear and the angular spectrum becomes more similar to the spectrum of the $\Lambda$CDM model. This is because, as mentioned earlier, then the suppression scale is not included in the observable window. In fact, when we have too many e-folds, we are shifting the observable window to a region where the PPS is quasi-invariant, resembling the PPS in the $\Lambda$CDM model. On the other hand, if a small number of e-folds is obtained during cosmic evolution, the observable window will cover a large interval of wave numbers with power suppression including the pivot scale, which will then not be placed in the quasi-invariant region of the PPS. 

One can see that, for the same number of e-folds, the PPS of the dressed metric model exhibits more suppression than the hybrid one. This is caused by the fact that the suppression scale in the PPS of the dressed metric approach is larger than its counterpart in the hybrid case. The multipole variation of this suppression is also slightly different in the two approaches. To elucidate if one of the two cases is observationally preferred, a Bayesian analysis should be conducted. This goes beyond the scope of this article, and will be the subject of future research.

\begin{figure}[h]
    \centering
    \includegraphics[width=1\textwidth]{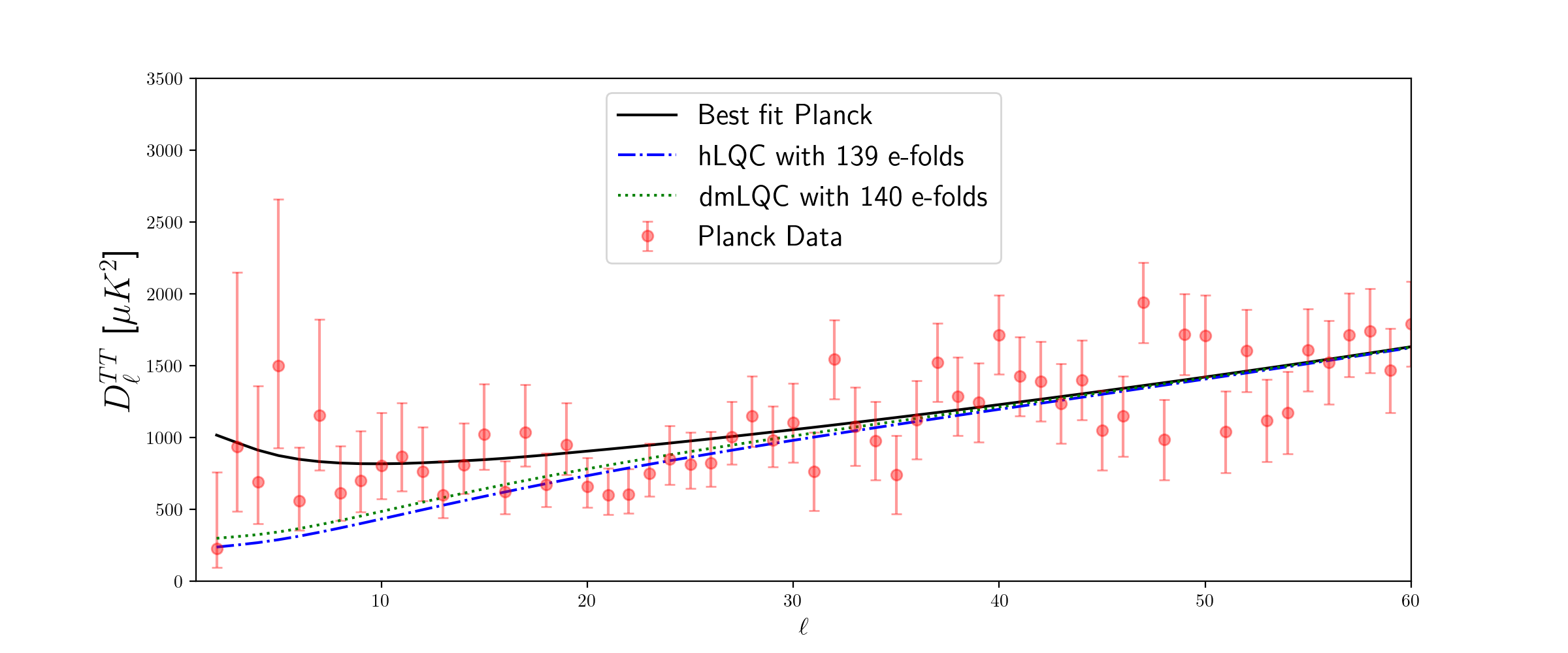} 
    \caption{Comparison between different temperature angular spectra, $D_{\ell}^{TT}$, obtained as explained in Figs. \ref{fig_APS_h} and \ref{fig_APS_dm} and restricted to the most interesting region of low multipoles with $\ell<60$. The solid dark line represents again the base-$\Lambda$CDM theoretical spectrum with the best fit to the Planck TT,TE,EE$+$lowE$+$lensing likelihoods of the Planck collaboration. The dot-dashed blue line and the dotted green line represent, respectively, the angular power spectrum for the hybrid approach to LQC (hLQC) with $139$ e-folds and the angular power spectrum for the dressed metric approach to LQC (dmLQC) with $140$ e-folds, both computed using the TT,TE,EE$+$lowE$+$lensing$+$BAO estimation of cosmological parameters by the Planck collaboration \cite{Planck_parameters}. For hybrid LQC, we have taken a scale factor at the end of the bounce period $a_0=1.347$ and a suppression scale $k_0=1.023$ in Planck units. For the dressed metric approach to LQC, we have set $a_0 = 1.674$ and $k_0=0.661$.}
    \label{fig_APS_compare_dm_h}
\end{figure}

From our computations, we see that the hybrid and dressed metric approaches lead to very similar angular spectra, even more if we compare the hybrid result for a given number of e-folds with the dressed metric spectrum with one more e-fold, as in the case shown in Fig. \ref{fig_APS_compare_dm_h}. The statistical significance of the possible improvements in the best fit to the observed data should be studied by Bayesian methods, as we have already mentioned. Similarly, in our calculation of the angular power spectra, we have fixed the values of the scale factor at the end of the bounce epoch, $a_0$, and the suppression scale, $k_0$, thereby fixing the impact of the quantum bounce. It would be most interesting to complete a Bayesian analysis in which these two quantities are viewed as phenomenological parameters. We emphasize that they completely characterize the power suppression in the PPS and therefore determine the suppression at low multipoles in the TT angular power spectrum, leaving clear imprints of the pre-inflationary LQC evolution and the choice of vacuum in the CMB. More precisely, these two parameters depend exclusively on the number of e-folds during the quantum bounce period, the number of e-folds during the subsequent cosmic evolution and the value of the energy density at the bounce. Changes in these quantities then imply a direct change in the PPS, and consequently a modification of the TT angular spectrum. 

Let us end this section by remarking that the quality of our parametrization demonstrates that, for phenomenologically interesting backgrounds and our choice of a vacuum, modifications of the power spectra of the standard model that could be produced by other scales of non-quantum origin can be ignored within our description. A more detailed analysis would be required for other cosmologies in which new physics is introduced in non-quantum regimes, in order to verify their impact on the power spectra. Within the considered context, our result is of utmost importance, as it confirms that quantum geometry corrections are more relevant than other pre-inflationary corrections (for our choice of vacuum state) and support the idea that any anomaly that could be detected in the CMB could be assigned to them. 

\section{Conclusions}

In this work, we have obtained a parametrization of the PPS for scalar and tensor perturbations in the hybrid and dressed metric approaches to LQC for a vacuum state selected with the NO-AHD proposal. This has also enabled us to compute the angular power spectrum of CMB temperature fluctuations and confront it with results of the Planck collaboration \cite{Planck_parameters}.

In both quantization approaches, we have found that the proposed parametrization of the PPS performs extremely well over a wide interval of wave numbers, including the suppression scale region and extending towards the ultraviolet region. This parametrization depends only on pre-inflationary physics and on the choice of vacuum that we have adopted. Our parametrization has, precisely, two new independent parameters compared to the familiar ones for the standard cosmological scenario with slow-roll inflation: these are the number of e-folds during the bounce period and the suppression scale. The latter can be expressed in terms of the total number of e-folds during cosmic evolution from the bounce to the present and of the maximum energy density, reached at the bounce. This critical density depends on the regularization of the Hamiltonian constraint in LQC, and also on constants of LQG such as the Immirzi parameter and the area gap. To reflect the possibilities that this dependence opens, we have treated it as a phenomenological parameter in the model studied in this work. On the other hand, our parametrization of the PPS can be directly interpreted as the usual PPS derived in the standard $\Lambda$CDM model with slow-roll inflation multiplied by a prefactor that accounts for the change of vacuum, from the Bunch-Davies state to the NO-AHD state, when all observable modes have frozen after exiting the cosmological horizon during the inflationary phase. 

From this parametrization of the PPS, for the two approaches to LQC studied in this work, we have computed the angular power spectrum of CMB temperature fluctuations and compared it to measurements by the Planck collaboration \cite{Planck_parameters}. We have checked that, for a not very large but phenomenologically acceptable number of e-folds from the bounce until today, around $138$--$143$ e-folds, the resulting angular spectrum shows an excellent agreement with the Planck measurements at high multipoles, indistinguishable from the best fit of the $\Lambda$CDM model. For a number of e-folds larger than $145$, the resulting angular spectrum is essentially indistinguishable from the standard angular spectrum of slow-roll inflation as, for all practical purposes, the cutoff scale lies outside the window of observable wave numbers. Furthermore, with our choice of vacuum state, the angular spectrum predicted by the LQC models investigated in this work, both in the hybrid and dressed metric approaches, exhibits a clear power suppression at low multipoles that may be key to decreasing the statistical significance of the CMB low-$\ell$ anomaly. In future research we want to carry out a Bayesian analysis to discuss the robustness and significance of this improvement. In particular, this analysis should contemplate variations of our two pre-inflationary parameters, which have not been explored in depth in our preliminary analysis here, as well as allow for
variation of the cosmological parameters of the $\Lambda$CDM model, $\{\Omega_bh^2,\Omega_ch^2, 100\theta_{\text{MC}}, \tau, \ln(10^{10} A_\text{s}), n_{\text{s}}\}$, for which we have used in this work their best fit values from TT,TE,EE$+$lowE$+$lensing$+$BAO (as reported in \cite{Planck_parameters}).

The novelty of this work with respect to other studies in LQC of the temperature angular spectrum (see e.g. Refs. \cite{ASr,ASr2,AgSr,AgSr2,hybridCMB,AAN,AM,waco2,Staro_LQC2,Staro_LQC3, Benito}) is evident from the following. First, we have introduced a new and manageable parametrization of the PPS which depends directly on the pre-inflationary physics and on the choice of the vacuum state. We would like to emphasize that the choice of vacuum plays a crucial role in this parametrization: adopting a different vacuum state would lead to entirely different results. This is because the vacuum determines the initial conditions for the evolution of the perturbations, thereby affecting the mode amplitudes when they freeze during the slow-roll regime. Consequently, a different choice of vacuum would yield a different PPS and, therefore, a different temperature angular power spectrum. Although several vacua have been studied in the literature in the context of LQC, leading to similar results for the temperature angular spectrum, their different properties induce at least differences in the value of the cutoff and in the slope of the PPS. Furthermore, for generic choices of vacuum state, the difference can be drastic, as we can understand by realizing that the change from Bunch-Davies (ignoring the pre-inflationary physics) to our vacuum state practically removes all the power for modes with wave number (sufficiently) below the effective cutoff. Another new result is the first confrontation in the literature of the NO-AHD vacuum predictions with CMB data showing a good agreement and the potential to alleviate the statistical significance of  CMB low-$\ell$ anomalies.

Let us finally remark that our parametrization and the analysis carried out in this work are valid for any inflaton potential assuming the approximations established in this work, so that the cosmic evolution is kinetically dominated from the bounce until the beginning of the slow-roll epoch. In this sense,the effect of the potential is negligible in the dynamics during the time that LQC effects are large, and hence our results can be extended to any allowed potential. Therefore, our parametrization provides a good framework to analyze whether a certain potential is favored or ruled out by observations, including those of the CMB and the late-time large-scale structure (see e.g.\ Refs. \cite{LSS, Chudaykin:2025vdh}), and whether the corresponding results may slightly differ from those for standard inflation.

In summary, the present study provides a solid foundation for investigating the imprints of LQC on the CMB and opens new avenues for future work in different directions. A detailed Bayesian analysis of cosmological data would be interesting to confront the model with observations. This analysis could constrain not only the standard cosmological parameters, but also in principle the fundamental quantities of the quantum model that are involved in the parametrization of the PPS and potentially alleviate anomalies and tensions present within and between cosmological datasets. From a theoretical perspective, another interesting topic of future research is whether the treatment of the background-dependent mass in the bounce period can be refined beyond the Pöschl–Teller approximation, determining how such refinements affect the resulting angular power spectrum (see footnote \ref{PTfoot}). This would provide the means to test the robustness of the approximation adopted for the bounce period in the present analysis.

\acknowledgments
The authors are grateful to K. Giesel and O. Friedrich for helpful discussions during the development of this project, as well as for carefully reading the manuscript and providing valuable suggestions for its improvement. They are also very thankful to B. Elizaga Navascu\'es for discussing fundamental ideas for this work and to J. Y\'ebana-Carrilero for helpful conversations. In addition, the authors would like thank Leonardo Ricci for conversations at the final stage of the project. This work was partially supported by MCIU/AEI/10.13019/501100011033 and FSE+ under the Grant No. PID2023-149018NB-C41. A.V.-B. acknowledges support of the WOST, WithOut SpaceTime project (https://withoutspacetime.org), supported by Grant ID\# 63683 from the John Templeton Foundation (JTF) (the opinions expressed in this work are those of the authors and do not necessarily reflect the views of the JTF).

\begin{appendix}

\section{Analytic solution of the mode equations}\label{Appendix_A}

\subsection{Hybrid approach}\label{Appendix_Hybrid}

In this Appendix, we summarize the most important details of the analytic solution to the mode equations in the hybrid approach. As mentioned in Sec. \ref{Hybrid quantization approach}, during the bounce period an analytic solution to these equations is not known for the exact background-dependent mass \eqref{eq_hyb_mass}. For this reason, an approximation to this mass is necessary. We use a Pöschl-Teller approximation, for which the general solution to the mode equations is \cite{NM} 
\begin{equation}
\mu_k^{(t,s)} =M_k\left [ x(1-x) \right ]^{-\frac{ik}{2\alpha} } {}_{2}F_1\left ( b_1^k,b_2^k; b_3^k;x \right ) 
+ N_k \left ( \frac{x}{1-x} \right )^{\frac{ik}{2\alpha}}{ }_{2}F_1\left ( b_1^k-b_3^k+1,b_2^k-b_3^k+1; 2-b_3^k;x \right ),
\end{equation}
where $x=\left [ 1+e^{-2\alpha \eta} \right ]^{-1}$ and the parameters $b^k_{j}$ ($j=1, 2, 3$) are given by
\begin{equation} 
    b_{1}^k+b_{2}^k+1=2 b_{3}^k ,\qquad b_{2}^k=\frac{1}{2}\left ( 1-\sqrt{1+\frac{32\pi \rho_c}{3\alpha ^2}} \right )-\frac{ik}{\alpha}, 
  \qquad   b_3^k =1-\frac{ik}{\alpha}.
\end{equation}
The NO-AHD vacuum is characterized by the choice of  mode constants $M_k=1/\sqrt{2k}$ and $N_k=0$ in the hybrid approach \cite{NM}\footnote{\label{foot}In the dressed metric case, $k$ must be replaced with $\bar{k} = \sqrt{k^2 + \bar{v}_0}$ \cite{AMV}.}. In terms of the scale factor at the end of the bounce period, $a_0$, and introducing the suppression scale $k_0 =a_0H_0$, we have
\begin{eqnarray}
    \notag b_{2}^k &=&\frac{1}{2}\Bigg[ 1-\sqrt{1+ \frac{4 (a_0^6-1)}{9         {\rm arcosh^2}(a_0^2)}       \left[{}_{2}F_1\left ( \frac{1}{6},\frac{1}{2};\frac{3}{2}; 1-a_0^6 \right)\right]^2 }  \Biggr]-i\frac{k}{k_0}  \frac{\sqrt{a_0^6-1}}{3 a_0^2 \  {\rm arcosh}(a_0^2)}   {}_{2}F_1\left ( \frac{1}{6},\frac{1}{2};\frac{3}{2}; 1-a_0^6 \right), \\ 
     b_3^k &=&1-i\frac{k}{k_0}  \frac{\sqrt{a_0^6-1}}{3 a_0^2 \   {\rm arcosh}(a_0^2)}   {}_{2}F_1\left ( \frac{1}{6},\frac{1}{2};\frac{3}{2}; 1-a_0^6 \right).     \label{eq_b_i}  
\end{eqnarray} 
Here, we have used that $\rho_c = 3k_0^2a_0^4/(8\pi)$ and the expressions of the conformal time, the scale factor, and the parameter $\alpha$ in the hybrid approach.

During the purely kinetic period, the general solution to the mode equations is \cite{NM}  
\begin{equation} \label{mu_k_kinetic_hybrid}
\mu_k^{(t,s)}= C_k \sqrt{\frac{\pi y}{4}} H_0^{(1)}(ky) + D_k \sqrt{\frac{\pi y}{4}}H_0^{(2)}(ky), 
\end{equation}
where $y= \eta -\eta_0 + 1/(2a_0H_0)$. The integration constants $C_{k}$ and $D_{k}$ are determined by imposing continuity of the modes up to first derivatives at the conformal time $\eta_0$, obtaining
\begin{eqnarray} \label{dressed_Ck_D_k}
\nonumber C_k &=& - \frac{i}{4}\sqrt{\frac{\pi k }{k_0}} \left(4a_0^4\right)^{\frac{ik}{2\alpha}}
\Biggl[\frac{\alpha b_M^{k}}{2a_0^4 k} H_0^{(2)}\left[\frac{k}{2k_0}\right]\; {}_{2}F_1\left( b_1^{k}+1, b_2^{k}+1; b_3^{k}+1  ;\frac{1+\sqrt{1-a_0^{-4}}}{2} \right)\\
&-&  \  \left\{\left( \frac{k_0}{k} -i\sqrt{1-a_0^{-4}} \right) H_0^{(2)}\left[\frac{k}{2k_0}\right] - H_1^{(2)}\left[\frac{k}{2k_0}\right]\right\} \; {}_{2}F_1\left(  b_1^{k}, b_2^{k}; b_3^{k}  ;\frac{1+\sqrt{1-a_0^{-4}}}{2} \right) \Biggr] ,\\
D_k &=& -  C_k\bigl |_{(2) \leftrightarrow (1) },
\end{eqnarray}
where $b_M^{k}=b_1^{k}b_2^{k}/b_3^{k}$ can also be expressed in terms of $a_0$ and $k_0$ using the above expressions, and the notation $(2) \leftrightarrow (1)$ indicates that the formula for $C_k$ is the same as for $D_k$ except for a flip in the kind of the Hankel functions, from the second kind to the first one.

During the inflationary period, we need to distinguish between the cases of scalar and tensor perturbations, because their background-dependent masses differ. Nonetheless, in the slow-roll regime, the only difference found in practice is the value of the slow-roll parameter $\nu$. For the tensor case we have $\nu_{(t)}= \sqrt{3 \epsilon_V+9/4 }$ and for the scalar case $\nu_{(s)} = \sqrt{ 9\epsilon_V - 3 \delta_V+9/4}$. The mass for the perturbations is then
\begin{eqnarray}
    s^{(t,s)}_{\text{SR}}=-\frac{\nu_{(t,s)}^2-\frac{1}{4}}{(\eta-\eta_e)^2}.
\end{eqnarray}
Recall that $\eta_e$ is the conformal time at the end of inflation. The general mode solution is given by
\begin{equation} \label{mu_k_SR_hybrid}
\mu_k^{(t,s)}= A_k^{(t,s)} \sqrt{\frac{\pi}{4}(\eta_e-\eta)} \, H_{\nu_{(t,s)}}^{(1)}[k(\eta_e-\eta)] 
+ B_k^{(t,s)} \sqrt{\frac{\pi}{4}(\eta_e-\eta)} \,H_{\nu_{(t,s)}}^{(2)}[k(\eta_e-\eta)].
\end{equation}

The value of the integration constants $A_{k}$ and $B_{k}$ are determined by imposing continuity of the modes up to first derivatives with the solution in the kinetic period at conformal time $\eta_i$, yielding 
    \begin{eqnarray} \label{eq_Ak_hyb}
\nonumber A_k^{(t,s)} &=& -  \frac{\pi^{3/2}}{16} \sqrt{\frac{k^3}{2 k_0 k_e k_i}}\, H_{\nu_{(t,s)}}^{(2)}\left[\frac{k}{k_e}\right]\left (4a_0^4\right )^{\frac{ik}{2\alpha}} \Bigg[ H_0^{(1)}\left[\frac{k}{2k_0}\right] \\ 
\nonumber&\times&\Bigg\{\frac{\alpha b_{M}^{k}}{2a_0^4k} \,{}_{2}F_{1}\left( b_{1}^{k}+1, b_{2}^{k}+1, b_{3}^{k}+1; x_0 \right) -   \left (\frac{k_0}{k}-i \sqrt{1-a_0^{-4}}  -\frac{H_1^{(1)}\left[k/2k_0\right]}{H_0^{(1)}\left[k/2k_0\right]} \right ) \, {}_{2}F_{1}\left( b_{1}^{k}, b_{2}^{k}, b_{3}^{k}; x_0 \right)\Bigg\} \\
\nonumber&\times& \Bigg\{H_0^{(2)}\left[\frac{k}{2k_i}\right]
 \left ( \frac{k_i}{k}+\frac{k_e}{k} \left(\nu_{(t,s)} +1/2\right)-\frac{H_{\nu_{(t,s)} +1}^{(2)}\left[k/k_e\right]}{H_{\nu_{(t,s)}}^{(2)}\left[k/k_e\right]}  \right ) - H_1^{(2)}\left[\frac{k}{2k_i}\right] \Bigg\} -   (1)\stackrel{0,1}\longleftrightarrow (2) \Bigg] ,\\
\end{eqnarray}
\begin{equation}\label{eq_Bk_hyb}
    B_k^{(t,s)} = -  A_k^{(t,s)}\bigl|_{(2)\; \stackrel{\nu,\nu+1}\longleftrightarrow \;(1) } \;.
\end{equation}
We have used again a compact notation for the interchange of first and second kind Hankel functions, with the order of the functions affected by this operation appearing above the double arrow.
In addition, we have introduced two new scales. The first one is the purely kinetic scale defined as $k_i= 1/(2y_i) = a_0^3H_0/a_i^2$, where we have used the definition of the variable $y$ and the expression of the scale factor during the kinetic epoch \eqref{scale_factor_kinetic}. The second scale is the inflationary scale $k_e= 1/ (\eta_e -\eta_i) \approx a_iH_i$, where we have used the slow-roll approximation in Eq. \eqref{scale_factor_SR}. 

\subsection{Dressed metric approach} \label{Appendix_Dressed}

For the dressed metric approach, during the bounce period we can obtain an analytic solution to the mode equations using a similar Pöschl-Teller approximation but adding a constant, as in Eq. \eqref{PT_dress}. The corresponding general solution is
\begin{eqnarray}
\bar{\mu}_k^{(t,s)} = \bar{M}_{k}\left[ \bar{x} \left( 1-\bar{x}\right) \right]^{-\frac{i\bar{k}}{2\bar{\alpha}}} {}_{2}F_{1}\left( \bar{b}_{1}^{\bar{k}}, \bar{b}_{2}^{\bar{k}}, \bar{b}_{3}^{\bar{k}}; \bar{x} \right) + \bar{N}_{k}\left ( \frac{\bar{x}}{1-\bar{x}} \right )^{\frac{i\bar{k}}{2\bar{\alpha}}} {}_{2}F_{1}\left( \bar{b}_{1}^{\bar{k}} - \bar{b}_{3}^{\bar{k}} + 1, \bar{b}_{2}^{\bar{k}} - \bar{b}_{3}^{\bar{k}} + 1, 2 - \bar{b}_{3}^{\bar{k}}; \bar{x} \right),
\end{eqnarray}
where $\bar{x} = \left[ 1 + e^{-2\bar{\alpha} \eta} \right]^{-1}$, $\bar{k} = \sqrt{k^2 + \bar{v}_0}$, and 
\begin{equation}
    \bar{b}_{1}^{\bar{k}}+\bar{b}_{2}^{\bar{k}}+1=\bar{b}_{3}^{\bar{k}},\qquad \bar{b}_{2}^{\bar{k}}=\frac{1}{2}\left ( 1-\sqrt{1+\frac{4(\bar{U}_0 - \bar{v}_0)}{\bar{\alpha} ^2}} \right )-\frac{i\bar{k}}{ \bar{\alpha}}, 
  \qquad  \bar{b}_{3}^{\bar{k}} = 1 - \frac{i\bar{k}}{ \bar{\alpha}}.
\end{equation}
Here, $\bar{b}_j^k$ ($j=1, 2, 3$) and $\bar{x}_0$ depend (apart from $k$) on $\bar{U}_0$, $\bar{\alpha}$, and $\bar{v}_0$, which in turn vary with the scale factor at the end of the bounce period $a_0$ and the critical density (and hence with $a_0$ and $k_0$) by means of Eq \eqref{dressedrelations} and the relation of these two parameters with the conformal time given by Eqs. \eqref{scale_factor_LQC} and \eqref{confytime}. 

In the kinetic period, on the other hand, the mass of the perturbations is the same as in the hybrid case and as in GR. Therefore, the general solution to the mode equations is also the same as in the hybrid approach, see Eq. \eqref{mu_k_kinetic_hybrid}. However the specific solution for the NO-AHD vacuum differs between the hybrid and the dressed approaches, as we have explained (see footnote \ref{foot}). This solution is characterized by the mode constants
\begin{eqnarray} \label{dressed_Ck_D_k}
\nonumber \bar{C}_k &=& - \frac{i}{4}\sqrt{\frac{\pi k^2}{\bar{k}k_0}} \left(\bar{x}_0(1-\bar{x}_0)\right)
^{-\frac{i\bar{k}}{2\bar{\alpha}}}\Biggl[ 
\frac{2\bar{\alpha}\bar{b}^{\bar{k}}_M}{k} \bar{x}_0(1-\bar{x}_0)H_0^{(2)}\left[\frac{k}{2k_0}\right]\; {}_{2}F_1\left( \bar{b}_1^{\bar{k}}+1, \bar{b}_2^{\bar{k}}+1; \bar{b}_3^{\bar{k}}+1  ;\bar{x}_0 \right) \\ 
&-&  \ \left\{\left( i \frac{\bar{k}}{k} (1-2\bar{x}_0)  +\frac{k_0}{k}\right) H_0^{(2)}\left[\frac{k}{2k_0}\right] - H_1^{(2)}\left[\frac{k}{2k_0}\right]\right\} \; {}_{2}F_1\left(  \bar{b}_1^{\bar{k}}, \bar{b}_2^{\bar{k}}; \bar{b}_3^{\bar{k}}  ;\bar{x}_0 \right)\Biggr] ,\\
\bar{D}_k &=& -   \bar{C}_k \bigl|_{ (2) \leftrightarrow (1)}.
\end{eqnarray}

The analysis of the scalar and tensor perturbations in the final period, the inflationary stage, is also similar to that presented for the hybrid approach, leading to the same general solution for the modes as in Eq. \eqref{mu_k_SR_hybrid}. In the same way as before, we determine the constants $\bar{A}_k$ and $\bar{B}_k$ by continuity up to the first derivatives, starting with our choice of vacuum solution in the bounce period for the dressed metric approach. This calculation gives
\begin{eqnarray} \label{eq_Ak_dress}
\nonumber \bar{A}_k^{(t,s)} &=& - \frac{\pi^{3/2}}{16} \sqrt{\frac{k^4}{2\bar{k}k_0 k_ik_e}}\, H_{\nu_{(t,s)}}^{(2)}\left[\frac{k}{k_e}\right]\left [\bar{x}_0(1-\bar{x}_0)\right ]^{-\frac{i\bar{k}}{2\alpha}} \Bigg[ H_0^{(1)}\left[\frac{k}{2k_0}\right]\\ \nonumber&\times& \Bigg\{\frac{2\bar{\alpha} \bar{b}^{\bar{k}}_M}{k} \bar{x}_0(1-\bar{x}_0) \,{}_{2}F_{1}\left( \bar{b}_{1}^{\bar{k}}+1, \bar{b}_{2}^{\bar{k}}+1, \bar{b}_{3}^{\bar{k}}+1; \bar{x}_0 \right) - \left (i \frac{\bar{k}}{k}(1-2\bar{x}_0)+  \frac{k_0}{k}-\frac{H_1^{(1)}[k/(2k_0)]}{H_0^{(1)}[k/(2k_0)]} \right ) \, {}_{2}F_{1}\left( \bar{b}_{1}^{\bar{k}}, \bar{b}_{2}^{\bar{k}}, \bar{b}_{3}^{\bar{k}}; \bar{x}_0 \right) \Bigg\} \\
\nonumber&\times& \Bigg\{H_0^{(2)}\left[\frac{k}{2k_i}\right]  \left ( \frac{k_i}{k}+\frac{k_e}{k}\left(\nu_{(t,s)} +\frac{1}{2}\right)-\frac{H_{\nu_{(t,s)} +1}^{(2)}[k/k_e]}{H_{\nu_{(t,s)}}^{(2)}[k/k_e]}  \right ) - H_1^{(2)}\left[\frac{k}{2k_i}\right] \Bigg\} -  (1)\stackrel{0,1}\longleftrightarrow (2)  \Bigg] ,\\
    \end{eqnarray}
\begin{eqnarray}\label{eq_Bk_dress}
        \bar{B}_k^{(t,s)} &=& - \bar{A}_k^{(t,s)}\bigl|_{(2)\; \stackrel{\nu,\nu+1}\longleftrightarrow \; (1) } \,.
\end{eqnarray}

\subsection{Transformation matrix method}
\label{Appendix_Transformation}

Starting with the general solution to the mode equations in each of the considered periods, expressed as a linear combination of two independent solutions that we know analytically, and recalling the requirement of continuity up to the first derivative at the matching points between those periods, a useful method to calculate the solution determined by our vacuum conditions is to employ linear transformation matrices. In this way, different approximations for individual periods can be tested efficiently using an abstract code for the computations.

Let us rename here $M_k$ and $N_k$ as $A_k^{(0)}$ and $B_k^{(0)}$ for convenience, and denote the subsequent integration constants by $A_k^{(j)}$ and $B_k^{(j)}$. The general mode solution of the Mukhanov-Sasaki equation in one epoch (characterized with the label $j$) can be written in terms of two independent solutions $\alpha^{(j)}_k$ and $\beta^{(j)}_k$ as
\begin{equation}\label{eq_mode_sol}
\mu_k^{(j)}=A_k^{(j)}\alpha_k^{(j)}+B_k^{(j)}\beta_k^{(j)} .
\end{equation}
Imposing continuity up to the first derivative between the different epochs, each of them defined between the respective conformal times $\eta_{j-1}$ and $\eta_{j}$ (with $\eta_{0}=0$), requires
\begin{equation}\label{eq_mode_cont}
    \mu_k^{(j)}(\eta_{j})=\mu_k^{(j+1)}(\eta_{j}) , \quad
\mu_k^{(j)\,\prime}(\eta_{j})=\mu_k^{(j+1)\,\prime}(\eta_{j}),
\end{equation}
where we recall that the prime denotes the derivative with respect to the conformal time $\eta$.

On the other hand, the linearity of the mode equations imply that
\begin{equation}\label{eq_forward_mode_eq}
    \begin{pmatrix}
        \mu_k^{(j)}(\eta_{j})\\\ \mu_k^{(j)\,\prime}(\eta_{j})
    \end{pmatrix}=\mathbf{F}_k^{(j)}(\eta_{j})
    \begin{pmatrix}
        A_k^{(j)}\\B_k^{(j)}
    \end{pmatrix}
\end{equation}
for all $j$, where we have introduced the notion of a forward transformation matrix,
\begin{equation}\label{eq_forward_matrix}
    \mathbf{F}_k^{(j)}(\eta)=\begin{pmatrix}
        \alpha_k^{(j)}(\eta)&\beta_k^{(j)}(\eta)\\\alpha_k^{(j)\,\prime}(\eta)&\beta_k^{(j)\,\prime}(\eta)
    \end{pmatrix} ,
\end{equation}
which provides the values at each point of the period of the mode solutions and their derivatives. Similarly we can introduce the notion of the inverse, backward transformation matrix,
\begin{equation}\label{eq_backward_matrix}
    \mathbf{B}_k^{(j)}(\eta)=\mathbf{F}_k^{(j+1)}(\eta)^{-1}.
\end{equation}
Then, the continuity conditions on the solutions allow us to express $A_k^{(j+1)}$ and $B_k^{(j+1)}$ as follows:
\begin{equation}
    \begin{pmatrix}
        A_k^{(j+1)}\\B_k^{(j+1)}
    \end{pmatrix}=\mathbf{B}_k^{(j)}(\eta_{j})\begin{pmatrix}
        \mu_k^{(j)}(\eta_{j})\\ \mu_k^{(j)\,\prime}(\eta_{j})
    \end{pmatrix} .
\end{equation} 
When we combine the forward and backward transformation matrices, we obtain the transformation matrix
\begin{equation}\label{eq_transformation_matrix}
    \mathbf{T}_k^{(j)}=\mathbf{B}_k^{(j)}(\eta_{j})\mathbf{F}_k^{(j)}(\eta_j).
\end{equation} 
Using this, we can describe the scalar and tensor amplitudes needed for the computation of the PPS as
\begin{equation}
    \begin{pmatrix}
        A_k^{(t,s)}\\B_k^{(t,s)}
    \end{pmatrix}\equiv\begin{pmatrix}
        A_k^{(n)}\\B_k^{(n)}
    \end{pmatrix}=\prod_{j=0}^{n-1}T_k^{(j)}\begin{pmatrix}
        A_k^{(0)}\\B_k^{(0)}
    \end{pmatrix},
\end{equation}
where we have assumed a total number of $n$ epochs.

For the dressed metric approach described in Sec. \ref{Appendix_Dressed}, for example, our vacuum conditions lead to
\begin{equation}
    A_k^{(0)}=\bar{M}_k , \quad
    B_k^{(0)}=\bar{N}_k.
\end{equation} 
For the bounce period we use the Pöschl-Teller approximation explained in the main text. Then the components $\mathbf{F}_{pq}^{(0)}(\eta)$  ($p,q=1,2$) of the forward transformation matrix $\mathbf{F}^{(0)}(\eta)$ have the following expressions:
\begin{eqnarray}
    \mathbf{F}_{11}^{(0)}&=&\left[ \bar{x} \left( 1-\bar{x}\right) \right]^{-\frac{i\bar{k}}{2\bar{\alpha}}} {}_{2}F_{1}\left( \bar{b}_{1}^{\bar{k}}, \bar{b}_{2}^{\bar{k}}, \bar{b}_{3}^{\bar{k}}; \bar{x} \right) ,\\
    \mathbf{F}_{12}^{(0)}&=&\left ( \frac{\bar{x}}{1-\bar{x}} \right )^{\frac{i\bar{k}}{2\bar{\alpha}}} {}_{2}F_{1}\left( \bar{b}_{1}^{\bar{k}} - \bar{b}_{3}^{\bar{k}} + 1, \bar{b}_{2}^{\bar{k}} - \bar{b}_{3}^{\bar{k}} + 1, 2 - \bar{b}_{3}^{\bar{k}}; \bar{x} \right) ,\\
    \mathbf{F}_{21}^{(0)}&=&\left[ \bar{x} \left( 1-\bar{x}\right) \right]^{-\frac{i\bar{k}}{2\bar{\alpha}}}  \Bigg[ -i\bar{k}\left(1-2\bar{x}\right){}_{2}F_{1}\left( \bar{b}_{1}^{\bar{k}}, \bar{b}_{2}^{\bar{k}}, \bar{b}_{3}^{\bar{k}}; \bar{x} \right) \\
    \nonumber&+& 2\bar{\alpha}\left[ \bar{x} \left( 1-\bar{x}\right) \right]\left(\frac{\bar{b}_{1}^{\bar{k}}\bar{b}_{2}^{\bar{k}}}{\bar{b}_{3}^{\bar{k}}}\right){}_{2}F_{1}\left( \bar{b}_{1}^{\bar{k}}+1, \bar{b}_{2}^{\bar{k}}+1, \bar{b}_{3}^{\bar{k}}+1; \bar{x} \right) \Bigg]  ,\\
    \mathbf{F}_{22}^{(0)}&=&\left ( \frac{\bar{x}}{1-\bar{x}} \right )^{\frac{i\bar{k}}{2\bar{\alpha}}} \Big(i\bar{k}{}_{2}F_{1}\left( \bar{b}_{1}^{\bar{k}} - \bar{b}_{3}^{\bar{k}} + 1, \bar{b}_{2}^{\bar{k}} - \bar{b}_{3}^{\bar{k}} + 1, 2 - \bar{b}_{3}^{\bar{k}}; \bar{x} \right)\\
    \nonumber&+& 2\bar{\alpha}\left[ \bar{x} \left( 1-\bar{x}\right) \right]\left(\frac{(\bar{b}_{1}^{\bar{k}} - \bar{b}_{3}^{\bar{k}} + 1)(\bar{b}_{2}^{\bar{k}} - \bar{b}_{3}^{\bar{k}} + 1)}{2 - \bar{b}_{3}^{\bar{k}}}\right){}_{2}F_{1}\left( \bar{b}_{1}^{\bar{k}} - \bar{b}_{3}^{\bar{k}} + 2, \bar{b}_{2}^{\bar{k}} - \bar{b}_{3}^{\bar{k}} + 2, 3 - \bar{b}_{3}^{\bar{k}}; \bar{x} \right)\Big) ,
\end{eqnarray}
where we have employed the same definitions as in Sec. \ref{Appendix_Dressed}. \newline

On the other hand, in the kinetic period the mode constants can be encoded in the following forward transformation matrix:
\begin{eqnarray}
    \mathbf{F}_{11}^{(1)}&=&\sqrt{\frac{\pi}{8k_0}}H_0^{(1)}\left(\frac{k}{2k_0}\right) ,\\
     \mathbf{F}_{12}^{(1)}&=&\sqrt{\frac{\pi}{8k_0}}H_0^{(2)}\left(\frac{k}{2k_0}\right) ,\\
    \mathbf{F}_{21}^{(1)}&=&  \sqrt{\frac{\pi}{8k_0}} \left[k_0 H_0^{(1)}\left(\frac{k}{2k_0}\right)-kH_1^{(1)}\left(\frac{k}{2k_0}\right) \right] ,\\
    \mathbf{F}_{22}^{(1)}&=&  \sqrt{\frac{\pi}{8k_0}} \left[k_0 H_0^{(2)}\left(\frac{k}{2k_0}\right)-kH_1^{(2)}\left(\frac{k}{2k_0}\right) \right] .
\end{eqnarray}
Finally, using the mode solutions during the inflationary period, the corresponding forward matrix reads
\begin{eqnarray}
    \mathbf{F}_{11}^{(2)}&=&\sqrt{\frac{\pi}{4k_e}}H_{\nu_{(t,s)}}^{(1)}\left(\frac{k}{k_e}\right) ,\\
    \mathbf{F}_{12}^{(2)}&=&\sqrt{\frac{\pi}{4k_e}}H_{\nu_{(t,s)}}^{(2)}\left(\frac{k}{k_e}\right),\\
    \mathbf{F}_{21}^{(2)}&=&\sqrt{\frac{\pi}{4k_e}}\left[-\frac{(2\nu_{(t,s)}+1)k_e}{2} H_{\nu_{(t,s)}}^{(1)}\left(\frac{k}{k_e}\right)+kH_{\nu_{(t,s)}+1}^{(1)}\left(\frac{k}{k_e}\right)\right],\\
    \mathbf{F}_{22}^{(2)}&=&\sqrt{\frac{\pi}{4k_e}}\left[-\frac{(2\nu_{(t,s)}+1)k_e}{2} H_{\nu_{(t,s)}}^{(2)}\left(\frac{k}{k_e}\right)+kH_{\nu_{(t,s)}+1}^{(2)}\left(\frac{k}{k_e}\right)\right].
\end{eqnarray}

We obtain the backward matrices for all periods by calculating the inverse of the given forward matrices. After combining these results into the respective transformation matrices, we end up with
\begin{equation}
    \begin{pmatrix}
        \bar{A}_k^{(t,s)}\\\bar{B}_k^{(t,s)}
    \end{pmatrix}=T_k^{(2)}T_k^{(1)}T_k^{(0)}\begin{pmatrix}
        \bar{M}_k\\\bar{N}_k
    \end{pmatrix}.
\end{equation}
Using this method it is easy to change transformation matrices, for example if we adopt other approximations during the bounce epoch, or to include additional periods, as for instance in Ref. \cite{AMV}, where a constant effective mass period was considered between the bounce epoch and the kinetic period. 
\end{appendix}

\end{document}